\documentclass[usenatbib]{mn2e}

\usepackage[dvipdfm]{graphicx}
\usepackage{amssymb}
\usepackage{txfonts}
\usepackage{url}

\newcommand{\integral}{{\textit{INTEGRAL}}}
\newcommand{\xte}{{\textit{RXTE}}}

\newcommand{\sax}{{\textit{Beppo\-SAX}}}
\newcommand{\gro}{{\textit{CGRO}}}

\newcommand{\swift}{{\textit{Swift}}}

\newcommand{\msun}{{\rm M}_{\sun}}

\newcommand{\g}{$\gamma$}

\let\oldhat\hat

\renewcommand{\hat}[1]{\oldhat{\mathbf{#1}}}

\newbox\grsign \setbox\grsign=\hbox{$>$} \newdimen\grdimen \grdimen=\ht\grsign
\newbox\simlessbox \newbox\simgreatbox \newbox\simpropbox
\setbox\simgreatbox=\hbox{\raise.5ex\hbox{$>$}\llap
     {\lower.5ex\hbox{$\sim$}}}\ht1=\grdimen\dp1=0pt
\setbox\simlessbox=\hbox{\raise.5ex\hbox{$<$}\llap
     {\lower.5ex\hbox{$\sim$}}}\ht2=\grdimen\dp2=0pt
\setbox\simpropbox=\hbox{\raise.5ex\hbox{$\propto$}\llap
     {\lower.5ex\hbox{$\sim$}}}\ht2=\grdimen\dp2=0pt
\def\ga{\mathrel{\copy\simgreatbox}}
\def\la{\mathrel{\copy\simlessbox}}

\title[The radio/X-ray correlation in Cyg X-3]{The radio/X-ray correlation in Cyg X-3 and the nature of its\\ hard spectral state}

\author[A. A. Zdziarski et al.]
{Andrzej A. Zdziarski,$^{1}$\thanks{E-mail: aaz@camk.edu.pl} Alberto Segreto$^2$ and Guy G. Pooley$^3$\\
$^1$Centrum Astronomiczne im.\ M. Kopernika, Bartycka 18, PL-00-716 Warszawa, Poland\\
$^2$INAF - Istituto di Astrofisica Spaziale e Fisica Cosmica, Via U. La Malfa 153, I-90146 Palermo, Italy\\
$^3$Cavendish Laboratory, J. J. Thomson Avenue, Cambridge CB3 0HE, UK\\
}

\date{Accepted 2015 November 9.  Received 2015 November 7; in original form 2015 August 9}

\pagerange{\pageref{firstpage}--\pageref{lastpage}}
\pubyear{2016}

\begin{document}

\maketitle

\label{firstpage}

\begin{abstract}
We study the radio/X-ray correlation in Cyg X-3. It has been known that the soft and hard X-ray fluxes in the hard spectral state are correlated positively and negatively, respectively, with the radio flux. We show that this implies that the observed $\sim$1--100 keV flux (which is a fair approximation to the bolometric flux) is completely uncorrelated with the radio flux. We can recover a positive correlation (seen in other sources and expected theoretically) if the soft X-rays are strongly absorbed by a local medium. Then, however, the intrinsic X-ray spectrum of Cyg X-3 in its hard state becomes relatively soft, similar to that of an intermediate spectral state of black-hole binaries, but not to their true hard state. We also find the radio spectra in the hard state of Cyg X-3 are hard on average, and the flux distributions of the radio emission and soft X-rays can be described by sums of two log-normal functions. We compare Cyg X-3 with other X-ray binaries using colour-colour, colour-Eddington ratio and Eddington ratio-radio flux diagrams. We find Cyg X-3 to be spectrally most similar to GRS 1915+105, except that Cyg X-3 is substantially more radio loud, which appears to be due to its jet emission enhanced by interaction with the powerful stellar wind from the Wolf-Rayet donor.
\end{abstract}

\section{Introduction}
\label{intro}

\begin{figure*}
\centerline{\includegraphics[height=6.5cm]{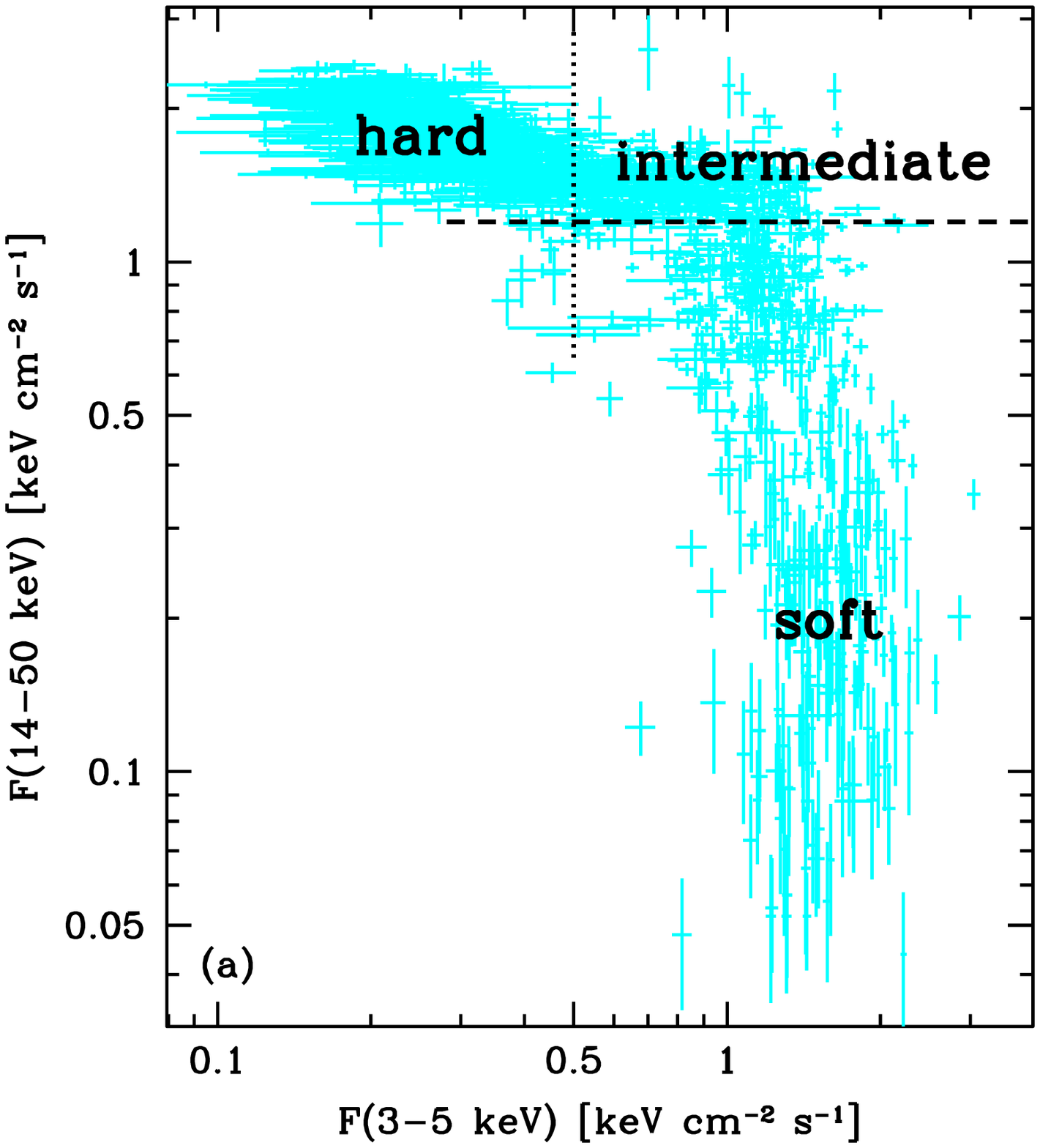}{\hskip 0.4cm}
\includegraphics[height=6.5cm]{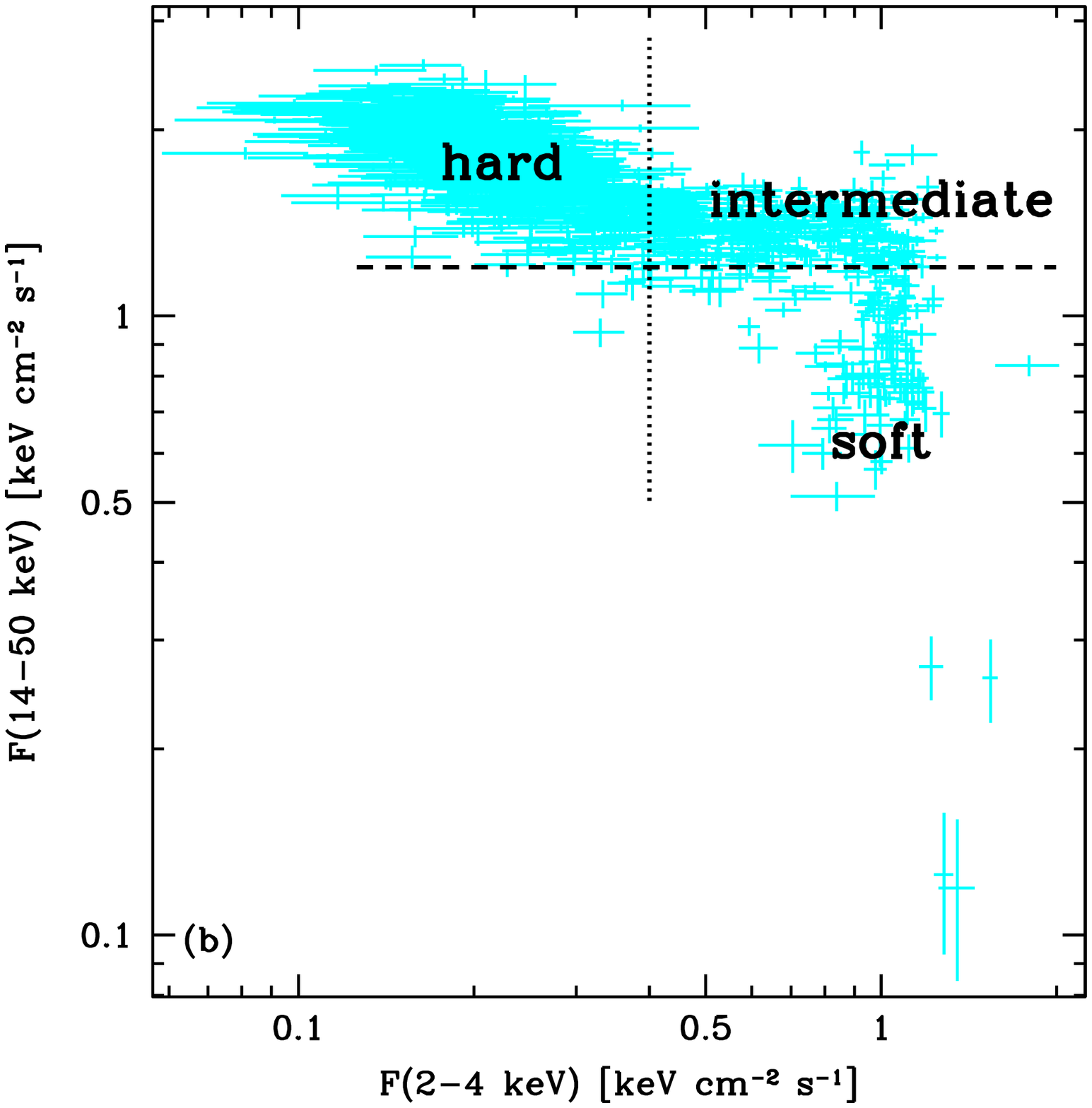}}
\caption{(a) The correlation between the daily-averaged energy fluxes in the 3--5 keV (ASM) and 14--50 keV (BAT) ranges. The hard and soft states are defined here by $F(3$--5 keV$)< 0.5$ keV cm$^{-2}$ s$^{-1}$ and $F(14$--50 keV$)< 1.2$ keV cm$^{-2}$ s$^{-1}$, respectively. These boundaries are marked by the dotted and dashed lines, respectively. An intermediate state corresponds to both the 3--5 keV and 14--50 keV fluxes above the respective boundaries. Only points with statistical significance $>3\sigma$ (for each of the fluxes) are shown. (b) The same except that the 2--4 keV flux from MAXI is used. The adopted hard-state condition (the dotted line) is here $F(2$--4 keV$)< 0.4$ keV cm$^{-2}$ s$^{-1}$.
}
\label{states}
\end{figure*}

Cyg X-3 is an binary system possessing a number of unique and puzzling characteristics. It is the brightest radio source among X-ray binaries \citep{mccollough99}, showing extremely strong radio outbursts and resolved jets (e.g., \citealt*{marti01,m01}). It is the only X-ray binary that is certainly powered by accretion for which high-energy \g-ray emission has been detected at a high statistical significance \citep{agile,fermi09}. It is the only known binary in the Galaxy containing both a compact object and a Wolf-Rayet star \citep*{v92,v96,v93,fender99}. Its orbital period of $P\simeq 0.2$ d is unusually short for a high-mass binary, which indicates a past spiral-in episode during a common-envelope evolutionary stage. Given the value of $P$ and estimates of the masses and mass-loss rate, the compact object orbits within the donor photosphere. 

In spite of its discovery as early as in 1966 \citep{giacconi67}, the nature of its compact object remains uncertain, due to the lack of fully reliable determination of the mass functions and inclination. Based on the currently available information, \citet*{zmb13} found the range of the compact-object mass of $2.4_{-1.1}^{+2.1} \msun$, allowing either a neutron star (NS) or a low-mass black hole (BH). The nature of the compact object is of significant interest, in particular because Cyg X-3 is a certain progenitor of a double compact system, after its Wolf-Rayet star explodes as a supernova \citep{belczynski13}. The presence of a BH over an NS is favoured by considering the X-ray and radio emission (\citealt*{h08,h09,sz08,szm08}, hereafter SZM08; \citealt{koljonen10}). \citet*{zmg10} have shown that the differences between the shape of the X-ray spectra of Cyg X-3 in its hard spectral state from those of confirmed BH binaries could be accounted for by Compton scattering in the strong stellar wind from the donor. That model also could account for the lack of high frequencies \citep*{alh09} in the power spectra of Cyg X-3. 

The distance to Cyg X-3 has been estimated as $D\simeq\! 7$--9 kpc \citep*{lzt09,d83,p00}: at this $D$, its observed (i.e., not corrected for intrinsic absorption) bolometric X-ray luminosity is $\sim 10^{38}$ erg s$^{-1}$. Its X-ray spectra have been classified into six states by \citet{koljonen10}, based on {\it Rossi X-ray Timing Explorer\/} (\xte) pointed observations. Here, we adopt a simpler classification into three states, hard, intermediate and soft, based on the relationship between the 3--5 keV and 14--50 keV fluxes. An unresolved issue is the value of the absorbing column within the stellar wind of the donor. For example, \citet{h08} found that the hard-state spectral data from \integral\/ can be fitted with either a high or low absorbing column. The two resulting models then significantly differ in both the spectral slope of the soft X-rays and in the bolometric flux. In the former and latter models, the intrinsic spectrum is soft, with the photon index of $\Gamma>2$, and hard, with $\Gamma<2$ (defined by the photon spectrum, $\propto E^{-\Gamma}$). Only the latter behaviour corresponds to the standard definition of the hard state, e.g., \citet{rm06}. 

Correlated radio X-ray behaviour of Cyg X-3 was discovered in the soft (1--6 keV) and hard (20--100 keV) X-rays by \citet{watanabe94} and \citet{mccollough99}, respectively. The correlations were then studied by \citet*{gfp03}, \citet{h08} and SZM08. Curiously, the soft X-rays are positively correlated with the radio flux in the hard state whereas the hard X-rays are anticorrelated. This differs, e.g., from another accreting high-mass binary, Cyg X-1, in which both the soft and hard X-rays are positively correlated with the radio emission in the hard state (\citealt{zspl11}; hereafter ZSPL11). This raises the issue of how the bolometric flux (dominated by the X-rays, emitted most likely by the accretion flow) is correlated with the radio emission, originating in the jets of this system. Theoretically, a positive correlation is expected \citep{hs03,mhd03} in a steady accretion-flow--jet system. Generally, this is because both the accretion luminosity and the jet power should be positively correlated with the accretion rate, and the radio flux is positively correlated with the jet power \citep{hs03}.

In this work, we study in detail the relationship between the radio emission and X-rays. In particular, we estimate the bolometric flux of the system, its relation to the radio flux, and the radio spectra. We then consider implications of our findings for the nature of this accreting source, and compare Cyg X-3 to other X-ray binaries.

\section{The radio and X-ray monitoring data}
\label{data}

We study here radio monitoring data at 15 GHz from the Ryle Telescope, which cover MJD 49231--53905, and the Arcminute Microkelvin Imager (AMI), MJD 54612--57211. The combined data set contains 76776 measurements. The AMI Large Array is the re-built and reconfigured Ryle Telescope. \citet{pf97} describe the normal operating mode for the Ryle telescope in the monitoring observations; the observing scheme for the AMI Large Array is very similar. The new correlator has a useful bandwidth of about 4 GHz (compared with 0.35 GHz for the Ryle), but the effective centre frequency is similar. The data are subject to variations in the flux calibration of $\la 10$ per cent from one day to another; we hereafter assume a 10 per cent fractional error. Also, we use the 2.25 GHz and 8.3 GHz monitoring data \citep{waltman96,waltman95,waltman96} from the Green Bank Interferometer (GBI)\footnote{\url{ftp://ftp.gb.nrao.edu/pub/fghigo/gbidata/gdata/}}. The data are for MJD 50409--51823 and contain 11022 measurements. 

We average all of the data used here over 1 day (except for the BATSE, see below). We have found that 1 day is a time scale which both strongly reduces the measurement errors and averages over the orbital modulation (with the orbital period of 1/5 d), see also \citet{z12b}. At the same time, it retains most of the correlated radio and X-ray variability. 

For soft and medium X-rays, we use daily-averaged monitoring data from the All-Sky Monitor (ASM; \citealt*{brs93,levine96}) on board \xte, giving count rates in three channels at energies of 1.3--3 keV, 3--5 keV and 5--12.2 keV, spanning MJD 50087--55915. We also use the data from the Monitor of All-sky X-ray Image (MAXI; \citealt{matsuoka09}) on board of {\it International Space Station\/} ({\it ISS}). We use daily averaged data during MJD 55058--57401. The data\footnote{\url{http://maxi.riken.jp/top/}} are given as photon fluxes in three channels, 2--4, 4--10 and 10--20\,keV. These are converted from the count rates by scaling to Crab assuming its photon spectrum of the form of 
\begin{equation}
F(E)/E=A \exp(-N_{\rm H}\sigma_{\rm H}) (E/1\,{\rm keV})^{-\Gamma},
\label{crab}
\end{equation}
where the H column density is $N_{\rm H}=3.5\times 10^{21}$ cm$^{-2}$, $\sigma_{\rm H}$ is the bound-free cross section per H atom averaged over the cosmic composition (calculated using {\tt wabs} in {\tt xspec}, \citealt{arnaud96}), $A=10$ cm$^{-2}$ s$^{-1}$, and $\Gamma=2.10$. We convert the MAXI data to the energy fluxes using the above. For other data sets used here, we use the above form of the Crab spectrum except for taking into account a break to $\Gamma=2.24$ at $\geq 100$ keV. 

For hard X-rays, we use daily-averaged data from the \swift\/ Burst Alert Telescope (BAT; \citealt{barthelmy05,m05,krimm13}) survey data. They were retrieved from the HEASARC public archive and processed using the {\tt BAT\_IMAGER} software \citep{segreto10}. They form a 14--195 keV 8-channel count-rate light curve, similar to that used in \citet{z12b}, but covering now a substantially longer interval of MJD 53348--57279. The energy fluxes for the ASM and BAT data have been obtained by using scaling to the Crab spectrum as given above. We note that \citet*{ijm07} have estimated that, if the photon index of the actual source differs from the Crab index by 1 or the column density is up to 10 times larger, the ASM energy-range flux differs by at most 30 per cent. We thus expect similarly minor effects on the scaling of the MAXI and BAT data.  

We also use data from the Burst and Transient Source Experiment (BATSE; \citealt{harmon02}) on board of {\it Compton Gamma Ray Observatory\/} (\gro). These monitoring data have been obtained using the Earth occultation analysis technique \citep{harmon02}. We use the same data as used by \citet{h08} and SZM08, which are given as 3-day average energy fluxes in the 20--50, 50--100 and 100--230 keV bands for the interval of MJD 48363--51686.

In Appendix \ref{lc}, we present and discuss the histograms of the flux distributions and the light curves for the X-ray monitors and for the Ryle/AMI.  

\section{The radio/X-ray correlation}
\label{rx}

\begin{figure*}
\centerline{\includegraphics[height=10cm]{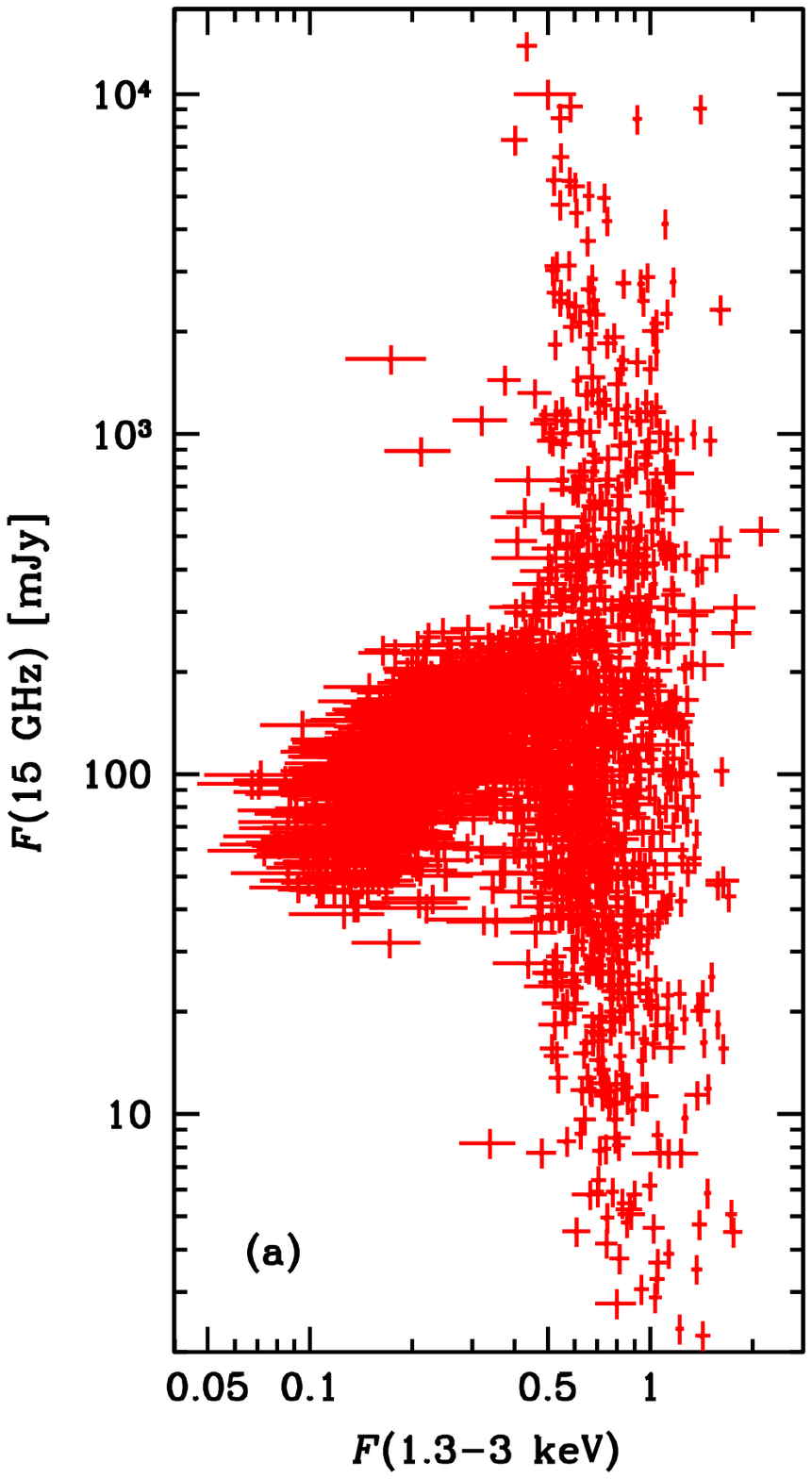}
\includegraphics[height=10cm]{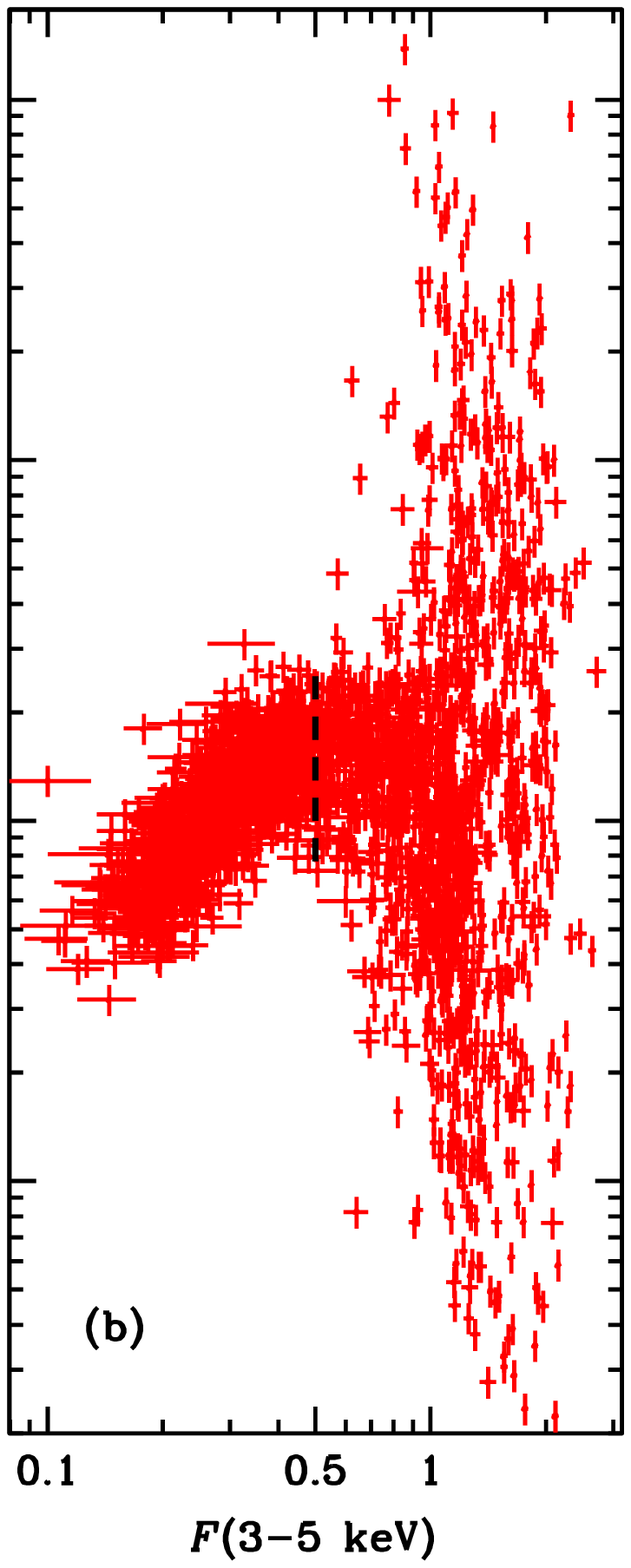}
\includegraphics[height=10cm]{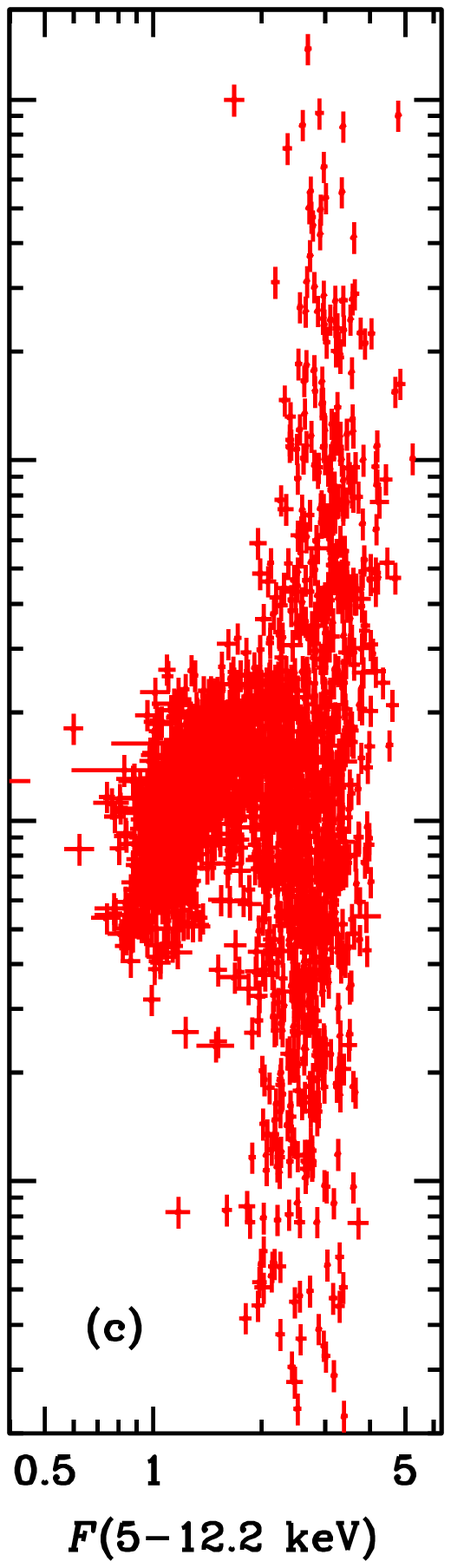}
\includegraphics[height=10cm]{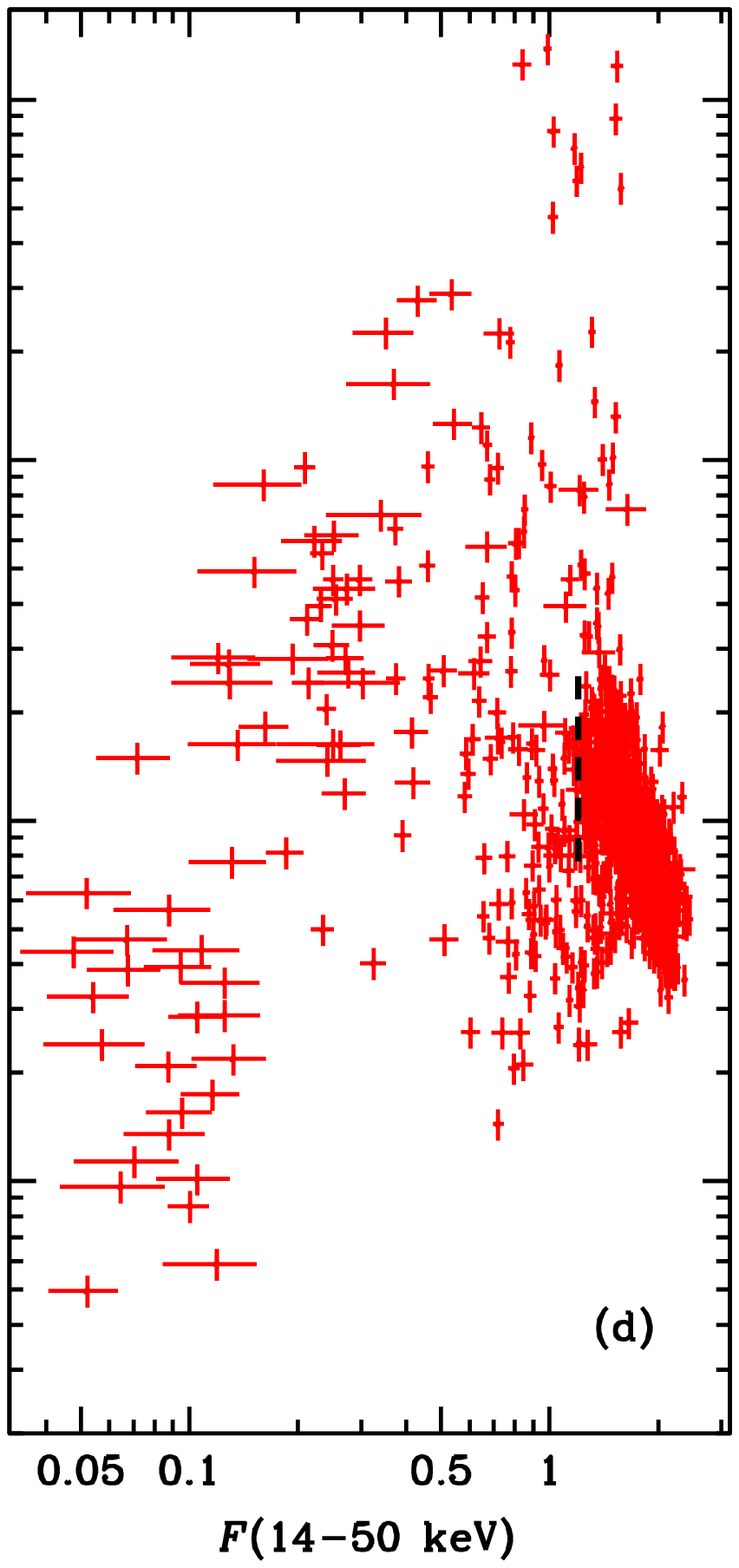}} {\vskip 0.4cm}
\centerline{\includegraphics[height=10cm]{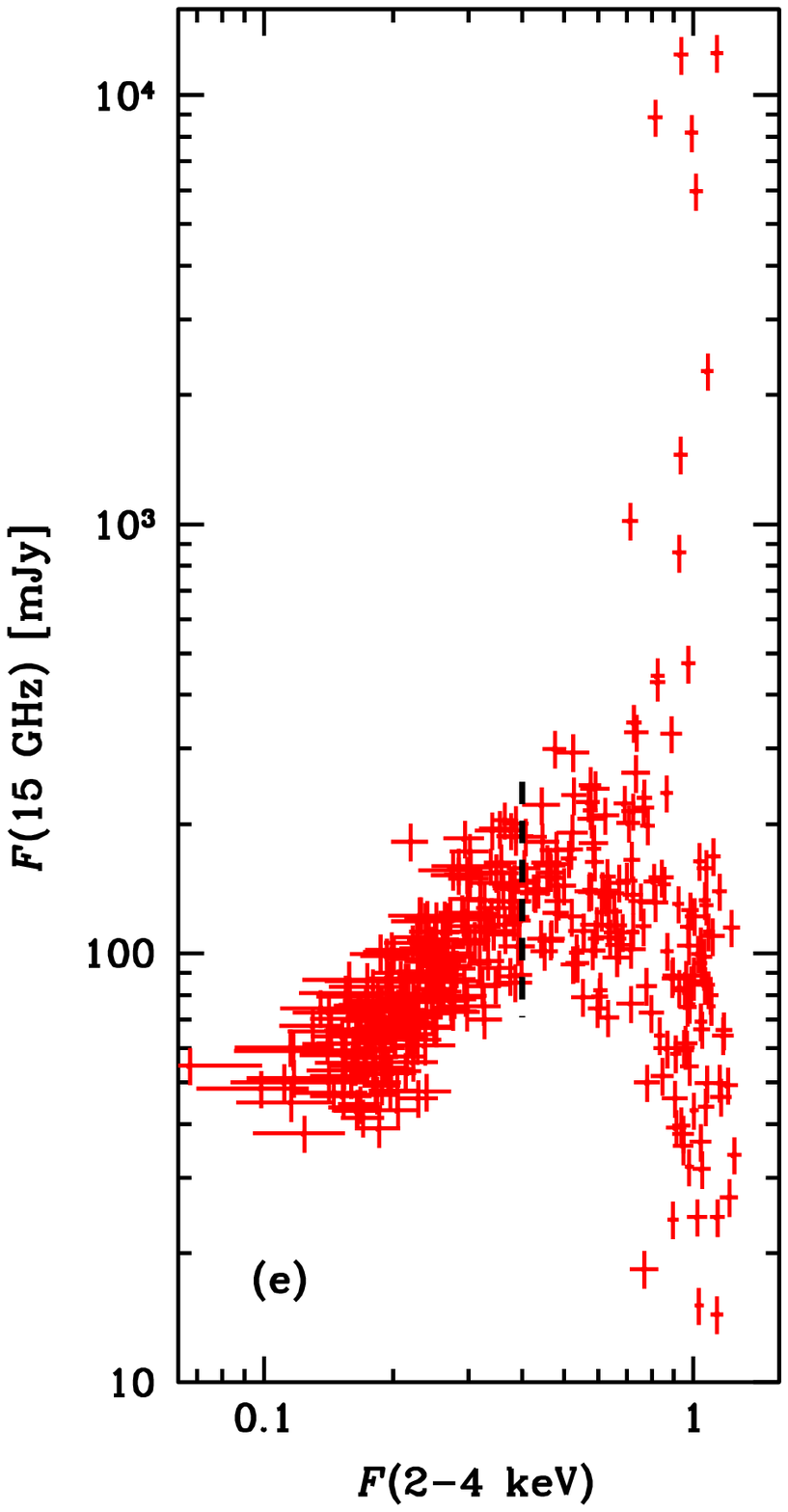}
\includegraphics[height=10cm]{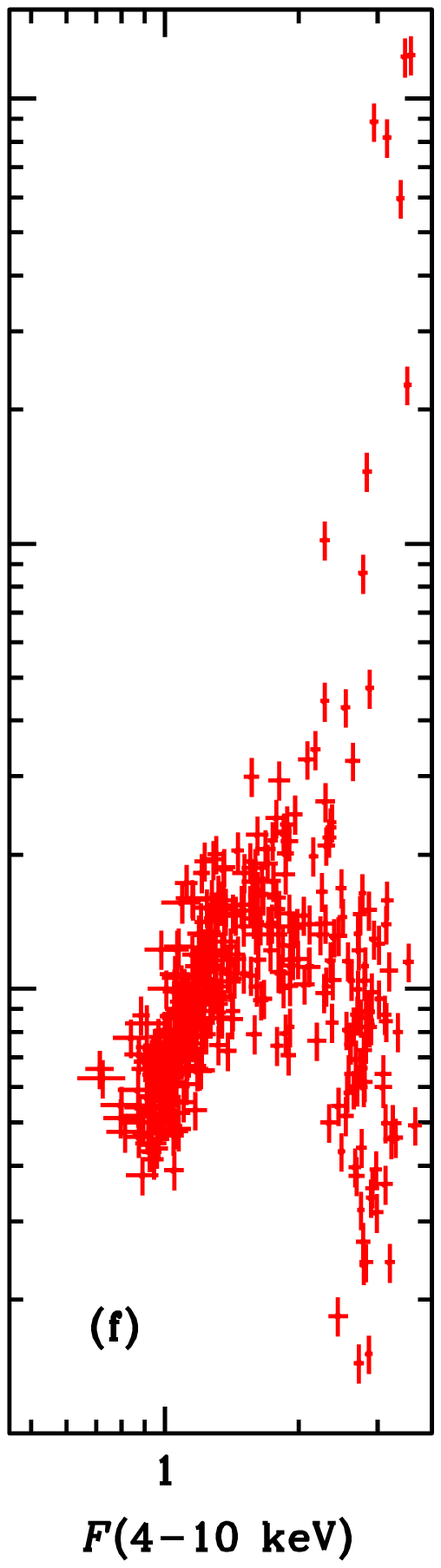}
\includegraphics[height=10cm]{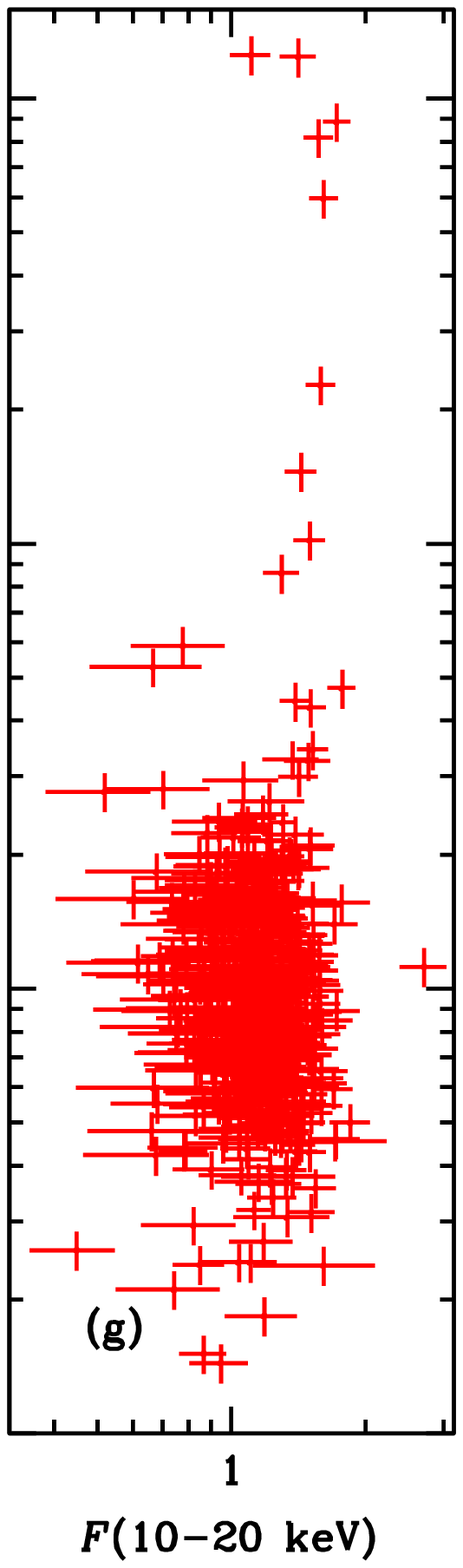}
\includegraphics[height=10cm]{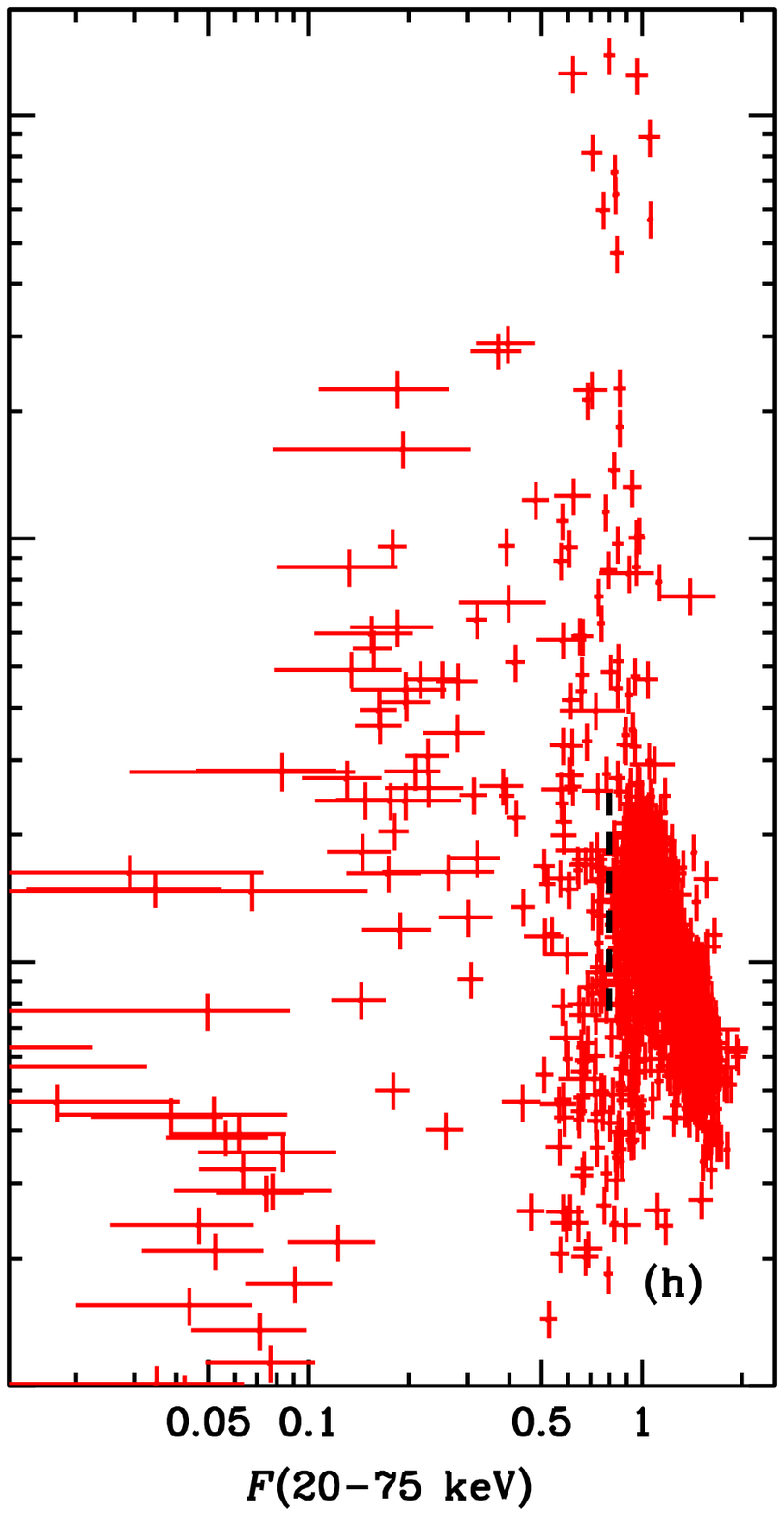}} 
\caption{The relationship between the 15 GHz flux and the X-ray flux in four energy bands based on (top panels) the ASM and BAT data and (bottom panels) the MAXI and BAT data, in units of keV cm$^{-2}$ s$^{-1}$. Only X-ray fluxes with significance $\geq 3\sigma$ are plotted. The dashed lines in (b) and (e) show the approximate upper boundary of the soft X-ray flux of the hard state, above which there are both the intermediate and soft states, see Fig.\ \ref{states}. The dashed lines in (d) and (h) show the approximate lower boundaries of the hard X-ray flux in the hard state, at 1.2 and 0.8 keV cm$^{-2}$ s$^{-1}$, respectively. The vertical extent of these lines is the same as in (b) and (e), and it corresponds to the range of the radio fluxes at the boundary of the hard state. We see the sign of the hard-state correlation changing around $\sim$10--20 keV. The strong radio flares occur only after the hard X-ray flux drops below its hard-state boundary (the dashed line), but then the radio flux in the soft state becomes (weakly) positively correlated with the hard X-rays. 
}
\label{rx4}
\end{figure*}

\begin{figure*}
\centerline{\includegraphics[height=10.cm]{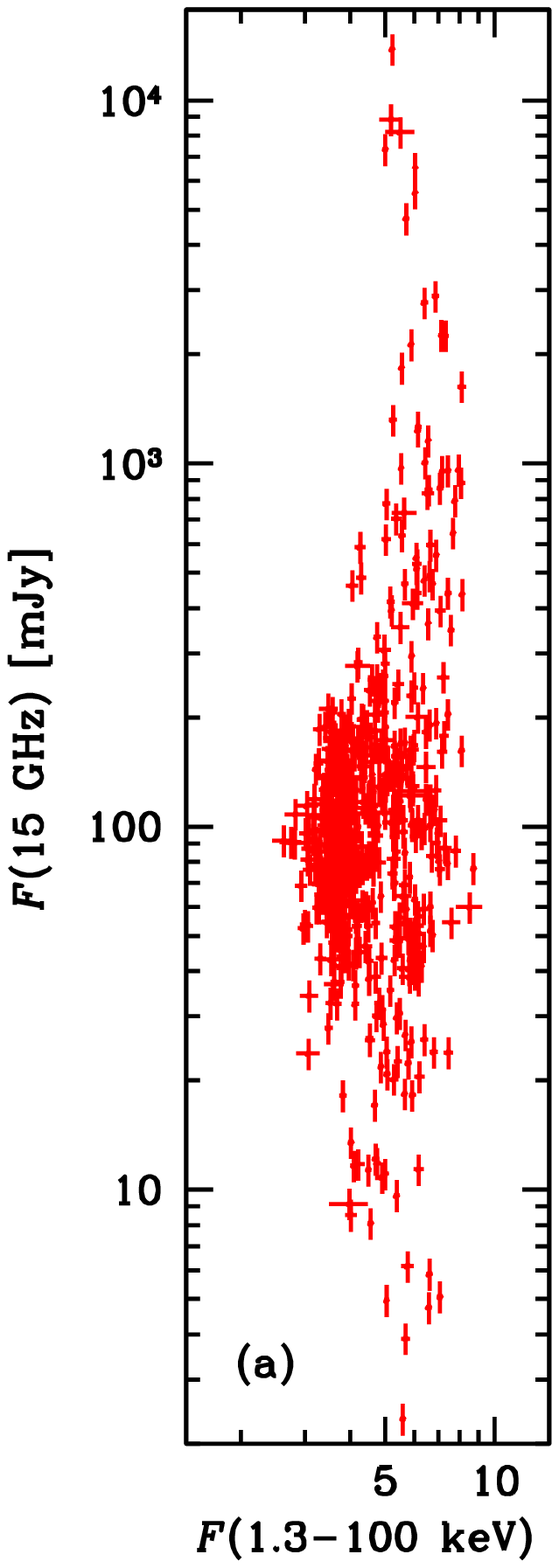}
\includegraphics[height=10.cm]{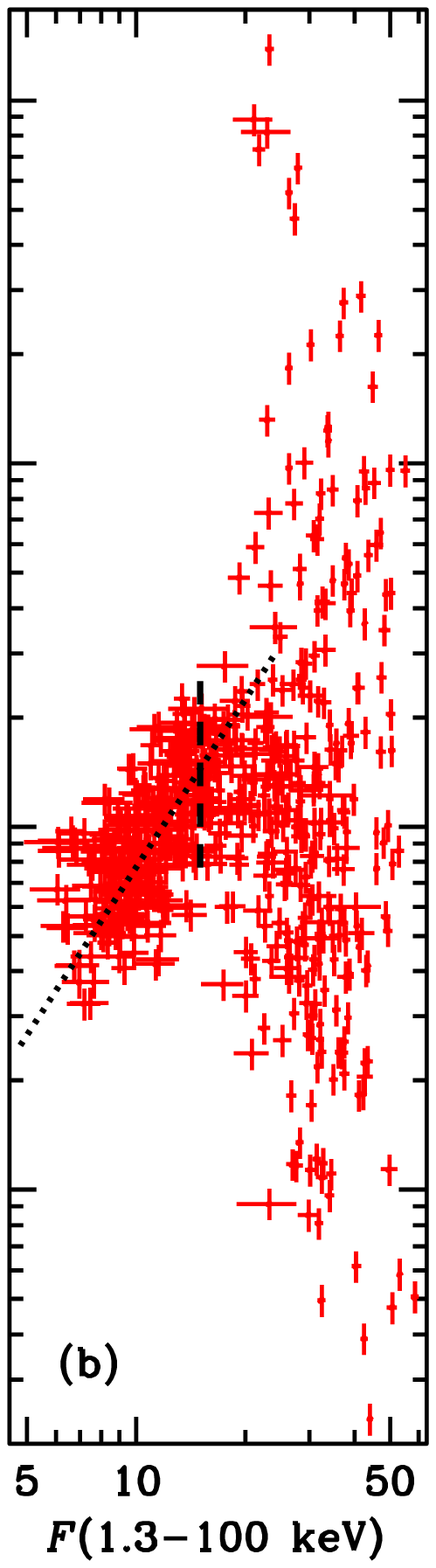}{\hskip 1cm}
\includegraphics[height=10.cm]{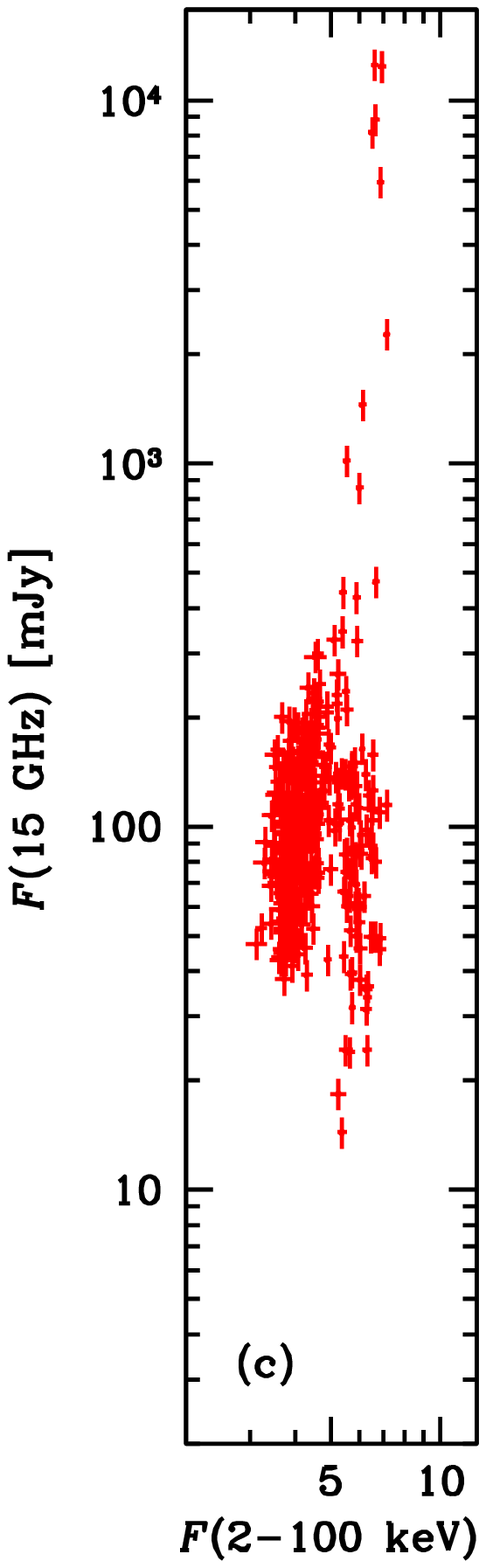}
\includegraphics[height=10.cm]{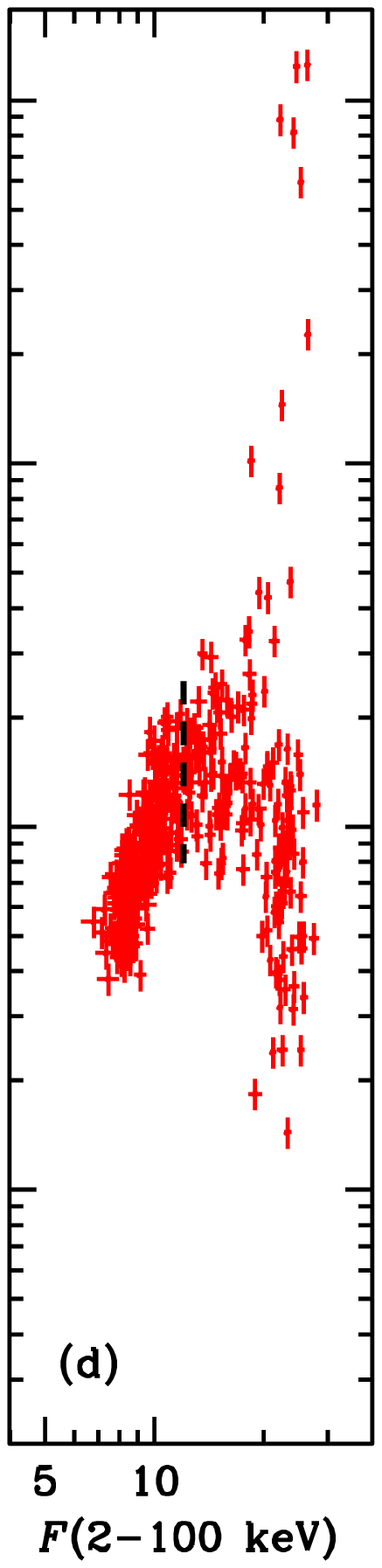}} 
\caption{Left panels (a--b): the relationship between the 15 GHz flux and the 1.3--100 keV flux, which is a good approximation to the bolometric flux. (a) The bolometric flux based on the ASM and BAT data as observed. (b) The flux corrected for absorption. Only X-ray fluxes with significance $\geq 5\sigma$ are plotted.The dotted line shows the power law fit to the hard state. Right panels (c--d): the same as (a--b) but based on the MAXI and BAT data.  The vertical dashed lines in (b) and (d) show the assumed upper boundary of the hard state at 15 and 12 keV cm$^{-2}$ s$^{-1}$, respectively. See Section \ref{rx} for details.
}
\label{rx_bol}
\end{figure*}

Fig.\ \ref{states} defines the spectral states of Cyg X-3 used in this paper via the relationship between the soft and hard X-rays, similarly to \citet{z12b}. The horizontal dashed line in both panels delineates the upper boundary of the soft state, with $F(14$--$50)\,{\rm keV})\leq 1.2$ keV cm$^{-2}$ s$^{-1}$. The vertical dotted lines delineate the upper boundary of the hard state, with $F(3$--$5)\,{\rm keV})\leq 0.5$ keV cm$^{-2}$ s$^{-1}$ for the ASM data, and $F(2$--$4)\,{\rm keV})\leq 0.4$ keV cm$^{-2}$ s$^{-1}$ for the MAXI data. The intermediate state is defined for the fluxes above both boundaries.

Fig.\ \ref{rx4} shows the energy dependence of the 15-GHz/X-ray relationship based on the ASM/BAT (top) and MAXI/BAT (bottom) data. Each data point corresponds to a daily average of the X-ray flux and the corresponding average radio flux. In Figs.\ \ref{rx4}(a--c) and (e--g), the data points showing a strong positive correlation seen on the left correspond to the hard state. The dashed lines in Figs.\ \ref{rx4}(b,e) show the upper flux limits of the hard state in the 3--5 and 2--4 keV bands, respectively, defined above. 

As found by SZM08, as the soft X-ray flux increases above the upper hard-state limit, the radio flux becomes highly variable, dropping to very low values and increasing to very high values, corresponding to major radio flares. Fig.\ \ref{rx4} does not show the temporal behaviour (shown in SZM08; see also Appendix \ref{lc}), but we do see the radio flux varying by 4 orders of magnitude while the soft X-ray flux changes only within a factor of $\sim$2.

As we also see in Fig.\ \ref{rx4}, the slope of the radio/X-ray correlation changes with the energy, becoming increasingly steep with increasing energy up to $\sim$10 keV, and then changing its sign to negative in the 14--50 and 20--75 keV bands. The vertical dashed lines in Figs.\ \ref{rx4}(d,h) show the lowest flux in those bands corresponding to the hard state. The change of the sign does happen between the ASM and BAT energy ranges, as we have checked by looking at the 14--20 keV band, which already shows a negative correlation (though with more scatter). Since that dependence is similar to those in the 14--50 and 20--75 keV bands, we do not show it here.

We have found the steepening of the correlation with the energy in the hard state in the simultaneous ASM/GBI data, at radio frequencies of 2.25 and 8.3 GHz, to be very similar to those shown in Figs.\ \ref{rx4}(a--c,d--g). Also, we have correlated the BATSE and 15 GHz data. We have found an anticorrelation at 20--100 keV range in the hard state similar to those shown in Figs.\ \ref{rx4}(d,h), but with much higher statistical scatter. The 100--230 keV BATSE data have then very limited statistics, given the spectrum of Cyg X-3 being strongly cut off at those energies. Given the similarities of those results to those using the ASM, BAT and MAXI and the 15-GHz data, we do not show them here.

\begin{figure*}
\centerline{\includegraphics[height=5.5cm]{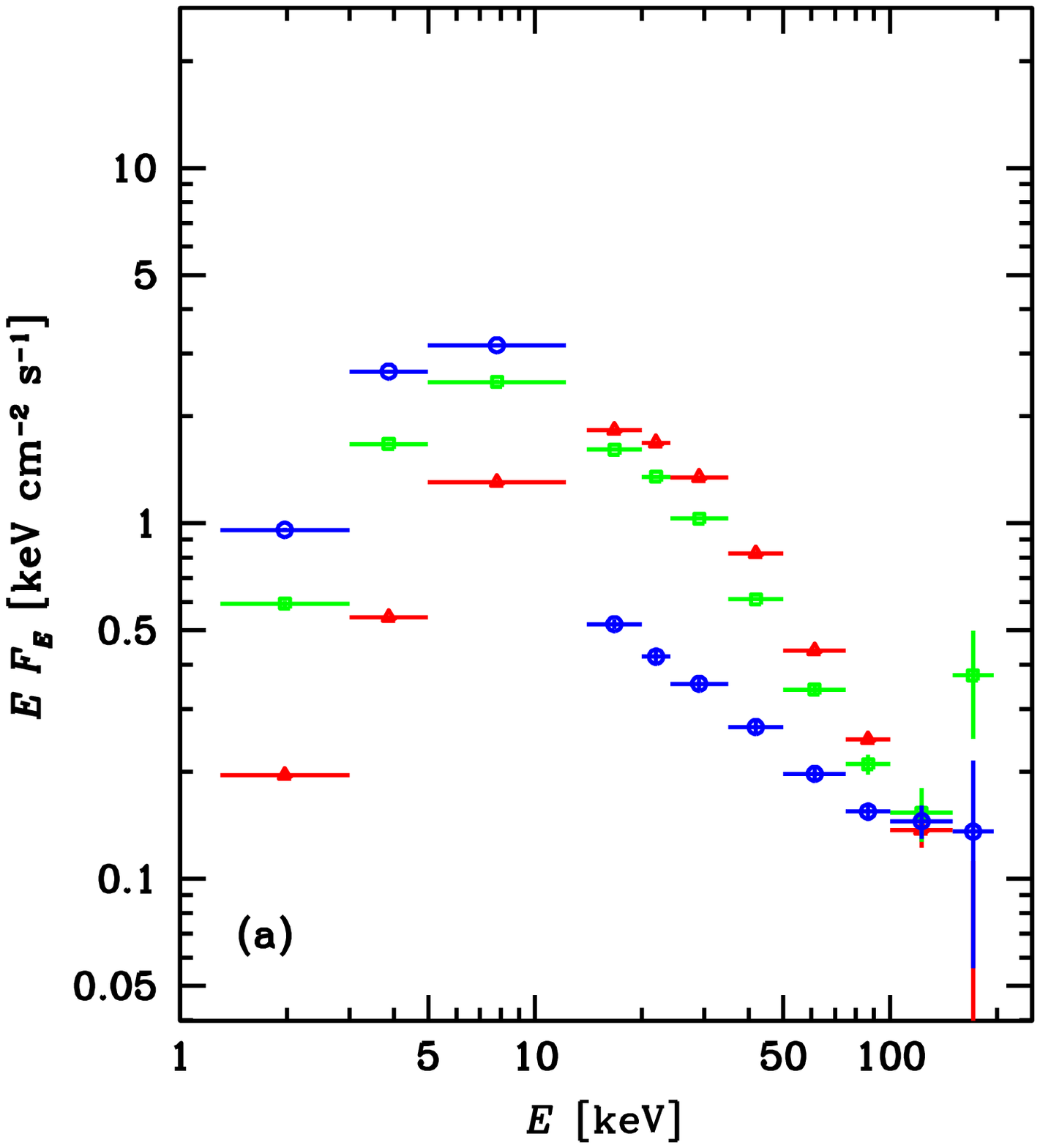}
\includegraphics[height=5.5cm]{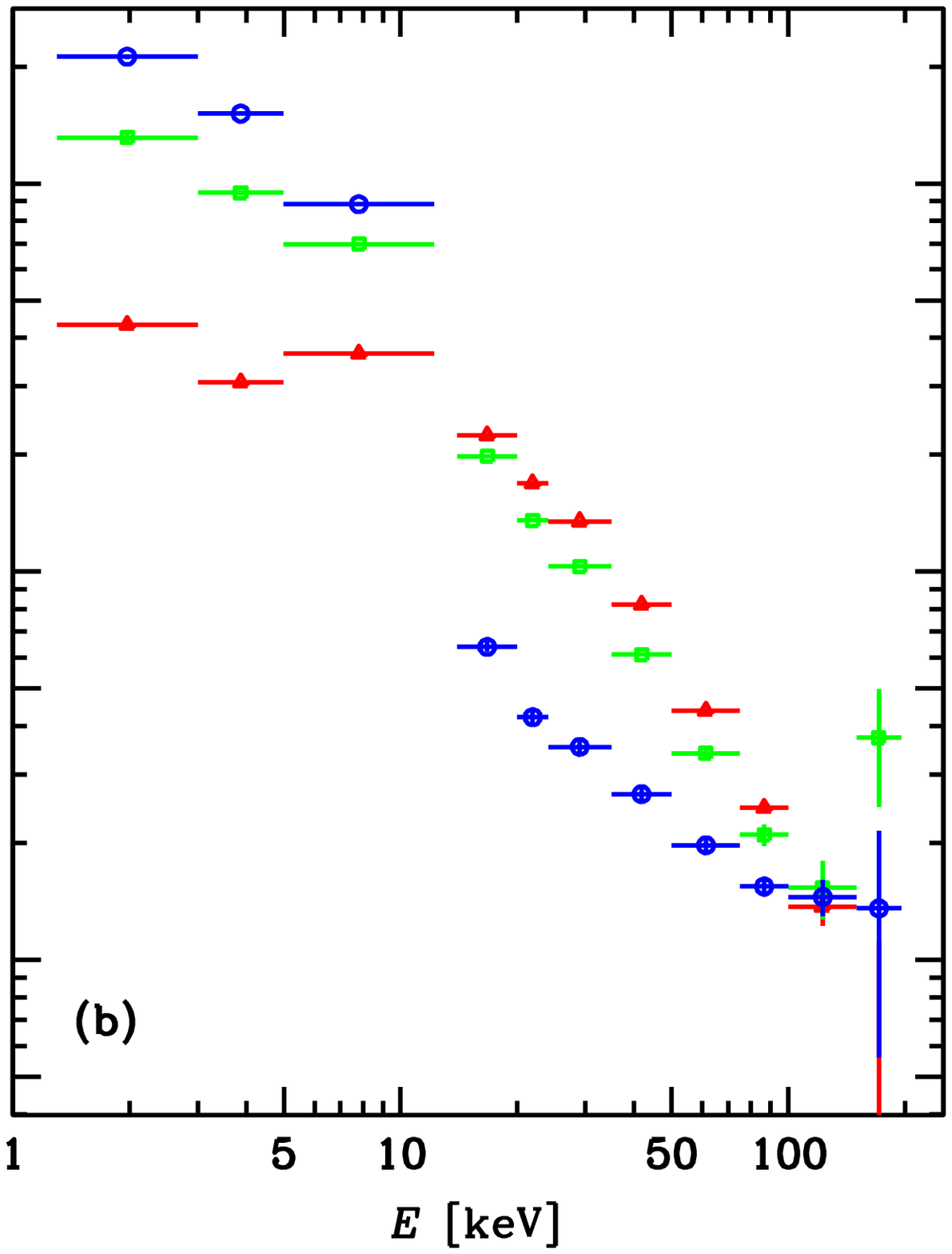}{\hskip 0.1cm}
\includegraphics[height=5.5cm]{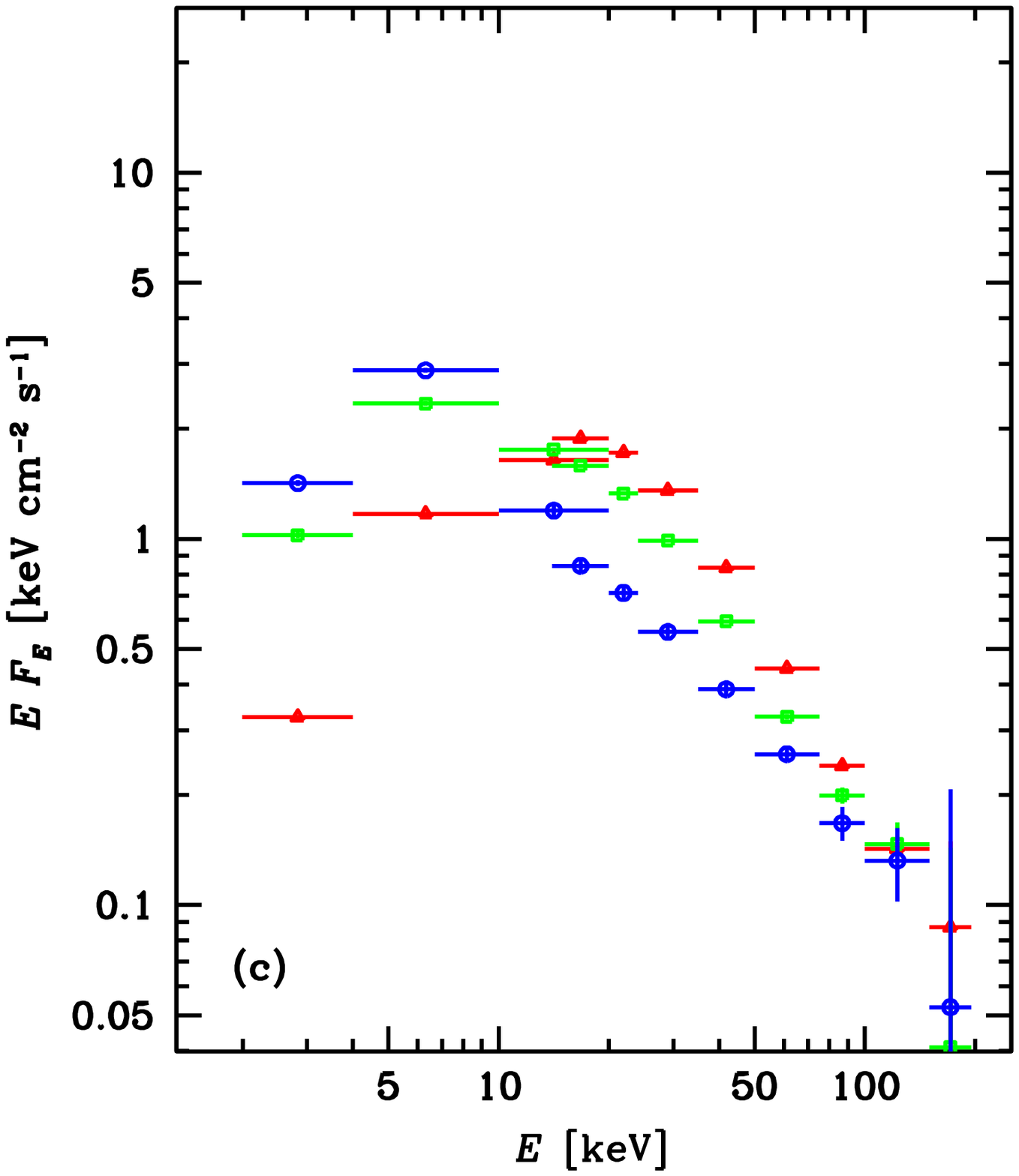}
\includegraphics[height=5.5cm]{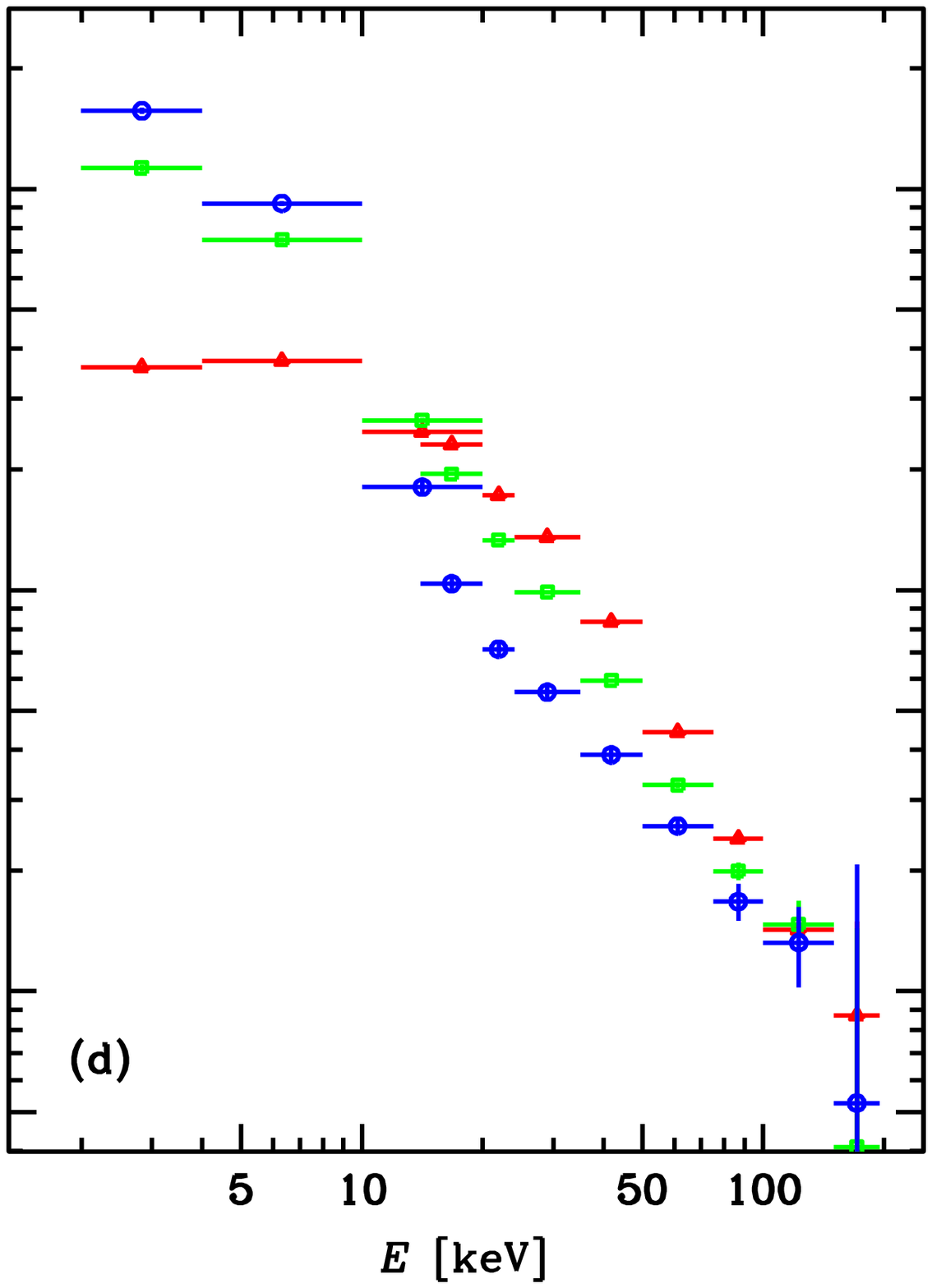}} 
\caption{Left panels (a--b): the average spectra based on the ASM and BAT data in the hard (red triangles), intermediate (green squares), and soft (blue circles) states; (a) observed, and (b) corrected for absorption. Left panels (c--d): the same but based on the MAXI and BAT data. See Section \ref{rx} for details.
}
\label{spectra}
\end{figure*}

The existence of the anticorrelation between radio and hard X-rays in the hard state raises the issue of how the bolometric luminosity is correlated with the radio flux. The observed $EF(E)$ spectra in the hard state peak at $\sim$20 keV (e.g., SZM08), and the spectral slope in soft X-rays is $\Gamma\sim 1$, which gives rise to the classification of this state as `hard'. Thus, the anticorrelation takes place in the spectral region in which at least $\sim$1/2 of the observed bolometric flux is emitted. 

Figs.\ \ref{rx_bol}(a,c) show the relationship between the 1.3--100 and 2--100 keV flux, respectively, and the radio flux. As it can be seen in the spectra of, e.g., SZM08, these X-ray energy ranges contain most (though not all) of the observed X-ray energy and thus they give fair approximations to the bolometric flux. (For the ASM/BAT data, we have neglected here a minor contribution from the 12.2--14 keV band, not covered by either ASM or BAT.) We see that there is hardly any correlation, with the data just forming a long vertical strip.

Obviously, there is substantial absorption in Cyg X-3, which needs to be taken into account. Its amount is, however, rather uncertain. \citet{h08} found that their hard-state data, from \integral, can be fitted by either weak absorption, with the resulting intrinsic slope below 10 keV of $\Gamma<2$, or strong one, with $\Gamma>2$. \citet{sz08} found that their hard-state \sax\/ data can be fitted with a weak absorption ($\Gamma<2$). As we find here, only relatively strong absorption, corresponding to $\Gamma\ga 2$, gives reasonable indices of the hard-state correlation of the intrinsic bolometric flux, $F_{\rm bol}$, with the radio flux, $F_{\rm R}$, $F_{\rm R}\propto F_{\rm bol}^p$. Thus, the hard state of Cyg X-3 would then be similar to an intermediate state of BH binaries.

The correlation index for most BH low-mass X-ray binaries (LMXBs) in the hard state for the 3--9 keV X-ray range is $p\simeq 0.6$--0.8 \citep{corbel00,corbel03,corbel04,corbel08,corbel13a,gfp03}. Then, some LMXBs have been found to have $p\simeq 1.4$, in particular the BH LMXB H1743-322 \citep{coriat11}, with the two branches intersecting at $L(1$--$10\,{\rm keV})\simeq 10^{38}$ erg s$^{-1}$, see fig.\ 9 in \citet{corbel13a}. As discussed in \citet{coriat11}, the two branches of the correlation correspond to the jet being powered by either radiatively inefficient ($p\simeq 0.7$) or radiatively efficient ($p\simeq 1.4$) accretion. 

Figs.\ \ref{rx_bol}(b,d) show the correlation between the radio flux and the intrinsic, unabsorbed, broad-band X-ray flux for a choice of the intrinsic absorber yielding $p\sim 1.5$. We use the ionized absorber model {\tt zxipcf} of \citet{reeves08} in {\tt xspec}, which is based on the treatment of ionized absorption by {\tt xstar} \citep{bk01}. It is characterized by a column density, $N$, and an ionization parameter, $\xi\equiv L_{\rm X}/(n r^2)$, where $n$ is the medium density and $r$ is the distance from the ionizing source to the medium. We assumed $N=10^{24}$ cm$^{-2}$ and $\xi= 10^{2.8}$ erg cm s$^{-1}$, which values are similar to those fitted in \citet{z12b} to the average \xte\/ X-ray spectra for the orbital phases around the superior and inferior conjunctions. In addition, we took into account interstellar absorption with $N_{\rm H}=1.5\times 10^{22}$ cm$^{-2}$ \citep{laugue72,cb73,dl90}. 

We have then created average X-ray spectra in the three spectral states, defined as in Fig.\ \ref{states}, based on the simultaneous ASM/BAT and MAXI/BAT data. The spectra as observed and the intrinsic ones, i.e., corrected for absorption, are shown in Fig.\ \ref{spectra}. We see that the intrinsic spectra in the hard state are almost flat in $E F(E)$ below $\sim$10 keV, with $\Gamma\sim 2$. 

The slope of the observed radio/X-ray correlation corresponding to the hard-state ASM/BAT spectrum shown in Fig.\ \ref{spectra}(b), and fitted as described in ZSPL11, is $p\simeq 1.5\pm 0.1$, which we show by the dotted line in Fig.\ \ref{rx_bol}(b). Given the overlap of the points in different states, the value of $p$ also bears a substantial systematic error. This slope is very similar to that found in the lower-branch (radiatively efficient) radio/X-ray correlation of LMXBs, see above. The corresponding slope for the MAXI/BAT data, shown in Fig.\ \ref{rx_bol}(d), is somewhat higher, which appears to be due to the lower boundary of the MAXI data being 2 keV compared to about 1.3 keV for the ASM data and the steepening of the correlation with the increasing energy up to $\sim$10 keV, found here.

\section{Radio spectra}
\label{radio}

\begin{figure}
\centerline{\includegraphics[width=5.8cm]{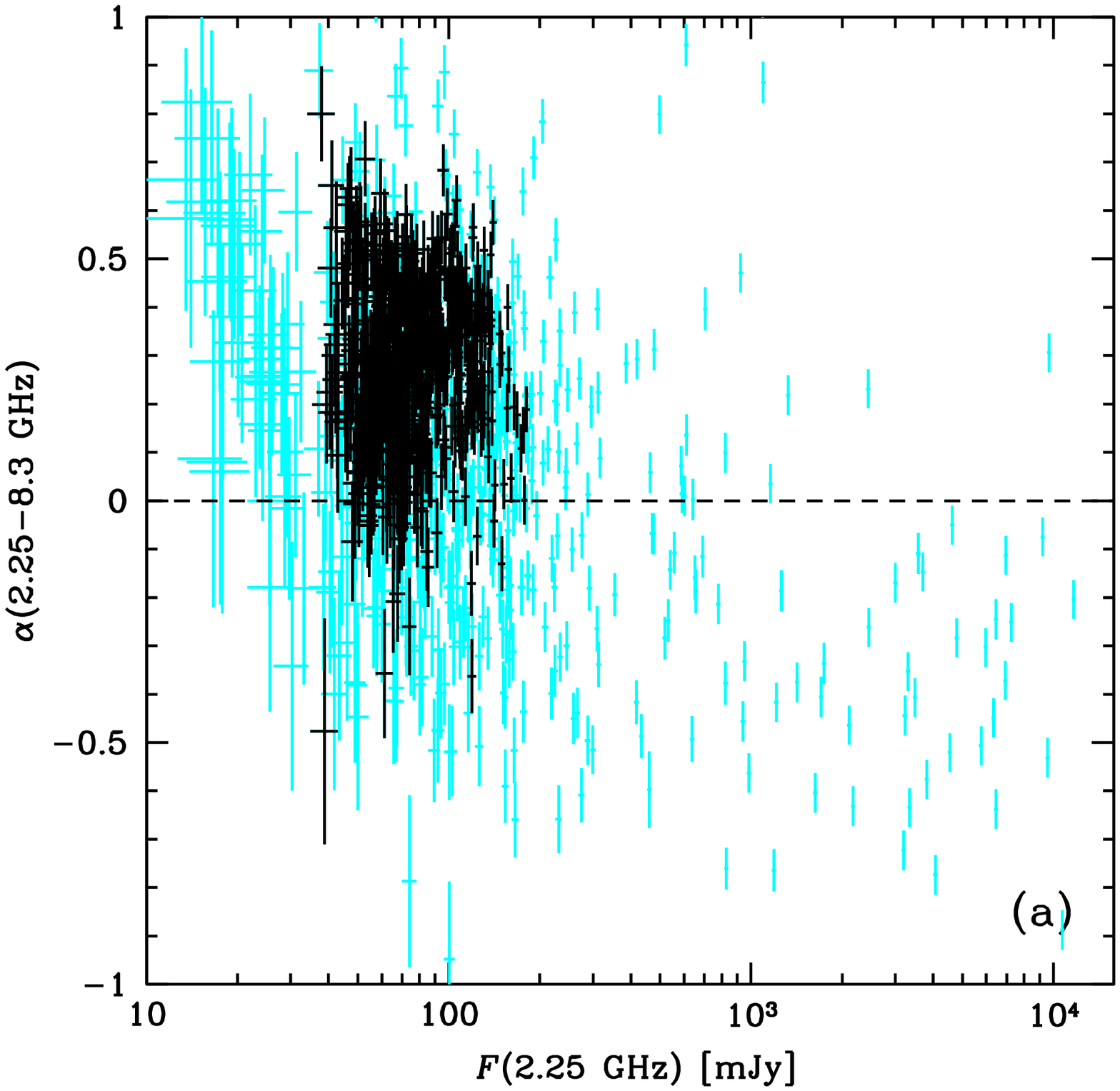}}
\centerline{\includegraphics[width=5.8cm]{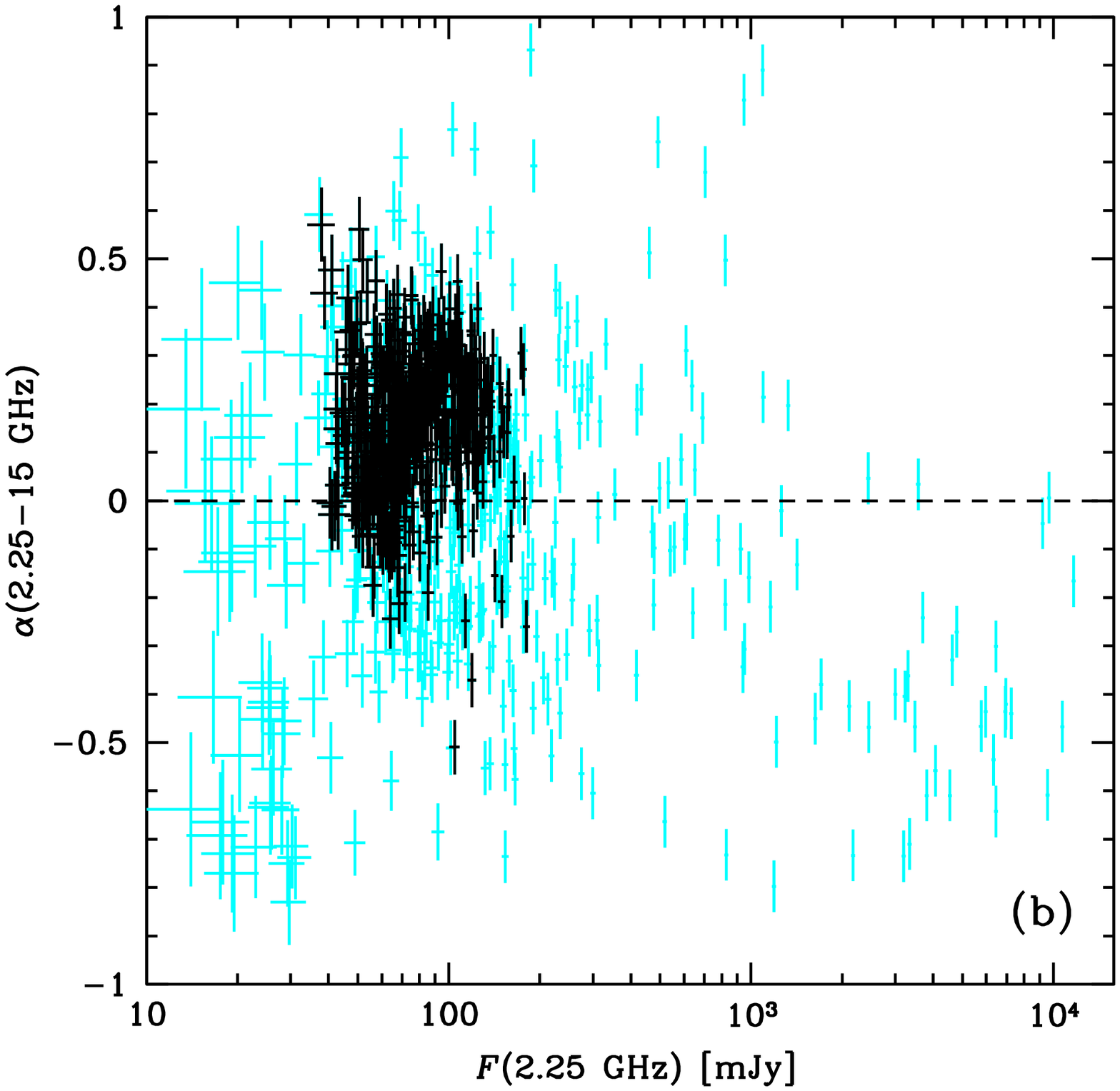}} 
\caption{The spectral index ($F_\nu\propto\nu^\alpha$) based on two radio frequencies vs.\ the flux of the lower frequency for 1-d average GBI and Ryle data. (a) 2.25--8.3 GHz; (b) 2.25--15 GHz. The hard state points identified by the simultaneous ASM data is shown by black symbols. Only the 2.25, 8.3 GHz and 3--5 keV flux measurements with the significance of $\geq 3\sigma$ have been taken into account. 
}
\label{alpha}
\end{figure}

In order to get more insight into the nature of the hard state, we consider the radio spectra based on three measurements, at 2.25, 8.3 and 15 GHz. We calculate the spectral index, $\alpha$, between two frequencies, defined by $F_\nu \propto \nu^{\alpha}$. We show the results in Fig.\ \ref{alpha}. We identify the hard state using the simultaneous ASM data, by our standard criterion of $F(3$--5 keV$)< 0.5$ keV cm$^{-2}$ s$^{-1}$, which we show by the black symbols. Note that some points not identified in this way due to the lack of simultaneous ASM data also belong to that state. We see that the hard state corresponds to the range of the 2.25-GHz fluxes of $\simeq$35--150 mJy and $\alpha\simeq -0.1$ to $+0.6$, with the average of $\langle\alpha\rangle\simeq 0.2$. These radio spectra are harder than those of Cyg X-1, with $\alpha\simeq 0$ \citep{fender00}. However, they are similar to those of those exhibited by some BH binaries.  For example, the LMXB GX 339--4 has shown $\alpha$ in the hard state during the 2010--11 outburst saturated at $\simeq 0.4$ (\citealt{corbel13b}, see their fig.\ 2). 

Thus, the hard state in Cyg X-3 is characterized by hard radio spectra, similar to the case of BH binaries. Such spectra are emitted by jets with the partially synchrotron self-absorbed emission \citep{bk79}. On the other hand, the soft state shows a wide range of $\alpha$, with majority of them with $\alpha$ between $-1$ and 0. The emission with $\alpha<0$ is likely optically thin synchrotron. However, the presence of spectra with $\alpha>0$ during the intermediate and soft states, see Fig.\ \ref{alpha}, indicates that synchrotron self-absorption is also taking place in those states.

\section{Discussion}
\label{discussion}

\subsection{The intrinsic radio/X-ray correlation index}
\label{intrinsic}

We have found, based on the ASM, MAXI, BAT and Ryle/AMI data, that the  Cyg X-3 radio flux in the hard state is highly variable and completely uncorrelated with the measured bolometric flux, as shown in Figs.\ \ref{rx_bol}(a,c). However, a positive correlation is recovered when taking into account absorption of soft X-rays provided it is strong enough, as shown in Figs.\ \ref{rx_bol}(b,d). We have found that an absorber similar to that fitted to the \xte\/ pointed data in \citet{z12b} gives a reasonable slope of the $F_{\rm bol}$-$F({\rm 15\,GHz})$ correlation, $p\simeq 1.5$, similar to those found in the lower branch of the LXMB radio/X-ray correlation \citep{corbel13b} and in Cyg X-1 in the hard state \citep{zdz12}. 

\begin{figure*}
\centerline{\includegraphics[height=8cm]{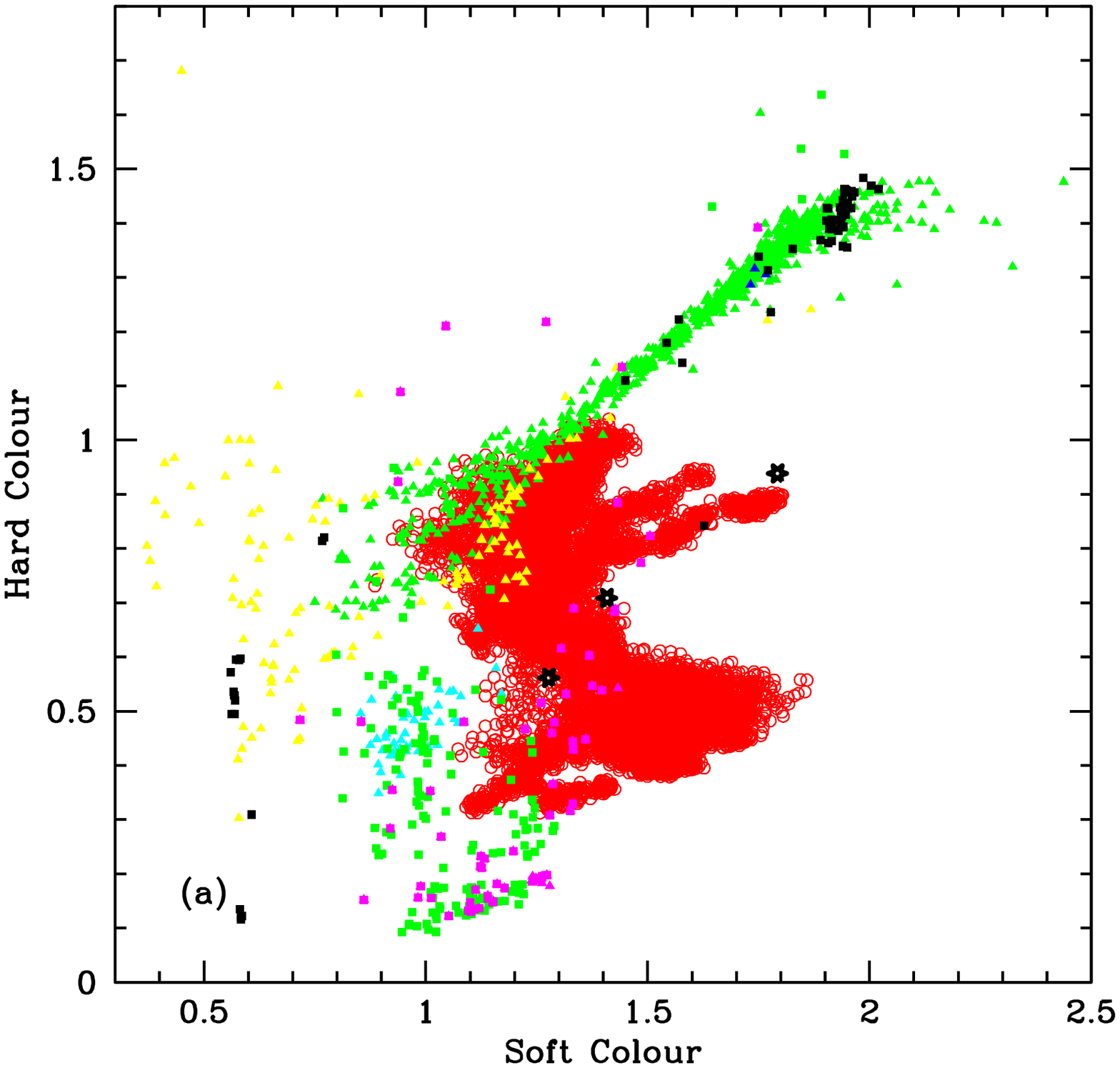}
\includegraphics[height=8cm]{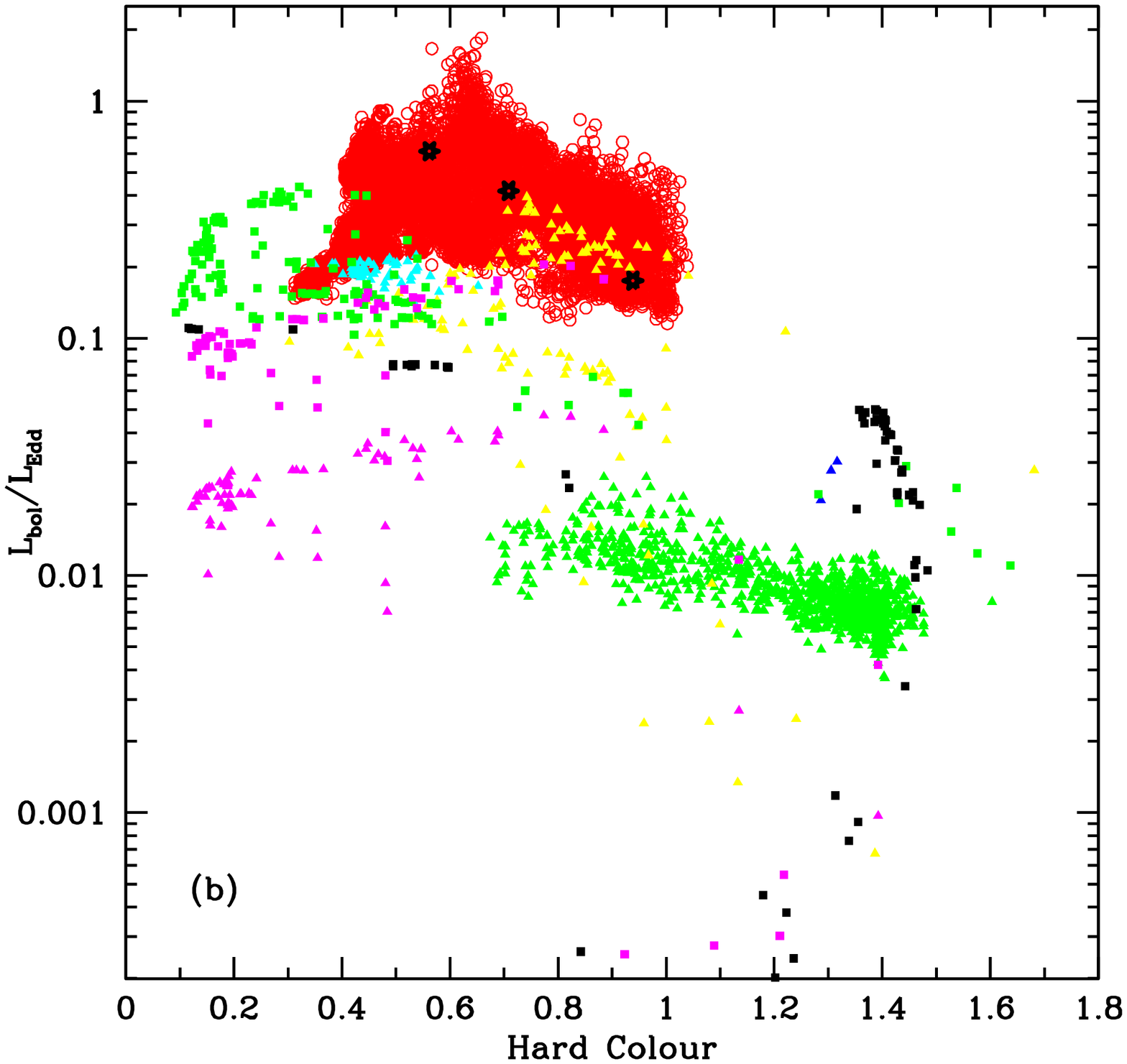}}
\caption{(a) The colour-colour diagram for selected Galactic BH binaries and Cyg X-3. The soft and hard colours are for the energy flux ratios of the bands 4--6.4 keV to 3--4 keV and 9.7--16 keV to 6.4--9.7 keV, respectively. (b) The Eddington ratio-hard colour diagram. On both panels, we show Cyg X-1 (green triangles, 1.86 kpc, $16\msun$), XTE J1650--500 (blue triangles, 2.6 kpc, $7\msun$), LMC X-1 (cyan triangles, 50 kpc, $10.9\msun$), XTE J1859+226 (yellow triangles, 8 kpc, $8\msun$), XTE J1118+480 (magenta triangles, 1.72 kpc, $7.5\msun$), LMC X-1 (green squares, 50 kpc, $10\msun$) GRO J1655--40 (magenta squares, 3.2 kpc, $6.0\msun$), GX 339--4 (black squares, 8 kpc, $12\msun$), GRS 1915+105 (red open circles, 8.3 kpc, $12.4\msun$) and Cyg X-3 (black stars, 8 kpc, $4\msun$). In parentheses, we give the plotted symbols and the adopted distance and mass. 
}
\label{lcd}
\end{figure*}

However, this index can be reduced due to an effect following from the physics of a continuous jet (e.g., \citealt{bk79,hs03}) coupled with the free-free absorption in the stellar wind. Namely, the height above the binary plane of a jet at which the bulk of partially synchrotron self-absorbed emission at a given $\nu$ emerges increases with the increasing $F_\nu$. This owes to the associated increase of the density of the jet relativistic electrons, which in turn increases the synchrotron self-absorption optical depth within the jet. For the standard model of \citet{bk79} and assuming that the radio emission is partially synchrotron self-absorbed (which leads to $\alpha\ga 0$), the predicted correlation between the height and the flux was derived in \citet{zdz12}, $z\propto F_{\rm intr}^q$, where $q=(s+6)/ (2 s+13)\simeq 1/2$. Here $s$ is the index of the power-law electron distribution in the jet, and $F_{\rm intr}$ is the emitted radio flux. As we have shown in Section \ref{radio}, the radio spectrum of Cyg X-3 in the hard state has indeed $\alpha\ga 0$, and the above formalism approximately applies. Then, at a higher flux, and thus a greater height, the photon path through the wind becomes shorter, and the orbit-averaged optical depth, $\bar{\tau}_{\rm ff}$, for free-free absorption becomes lower. A further effect is an increase of the temperature of the wind in response to the increased X-ray flux, which also reduces $\bar{\tau}_{\rm ff}$. The presence of these effects leads to the intrinsic correlation index, $p_{\rm intr}$, becoming lower than the one observed,
\begin{equation}
p_{\rm intr}= r p,\quad r= {{\rm d} \ln F_{\rm intr}\over {\rm d}\ln F_{\rm R}}
=1+{{\rm d}\bar{\tau}_{\rm ff} \over {\rm d}\ln F_{\rm R}},
\label{index}
\end{equation}
where $F_{\rm R}$ is the observed radio flux (ZSPL11). The value of $r$ can be estimated based on the dependence of the depth of an orbital modulation of the radio flux on $\nu$. In the case of Cyg X-1, \citet{zdz12} estimated $r\simeq 0.8$, leading to $p_{\rm intr}\simeq 1.4$ in that binary. 

In the case of Cyg X-3, the orbital modulation of the hard-state radio emission is very weak, with the full depth of the modulation of 15 GHz emission of $\simeq 5$ per cent (Zdziarski et al., in preparation). This is much weaker than that in the hard state of Cyg X-1, where it is $\simeq 30$ per cent \citep{zdz12}. This weakness is most likely due to two effects. First, the separation between the binary components in Cyg X-3 is about an order of magnitude lower than that of Cyg X-1 (given the orbital period of the former is much shorter than of the latter). Second, the height along the jet at which a given radio frequency is emitted is most likely higher in Cyg X-3 than in Cyg X-1, due to the accretion rate in the former being much higher than in the latter. Therefore, the asymmetry of the paths of 15-GHz photons through the wind between the superior and inferior conjunction in Cyg X-3 is much lower than in Cyg X-1. This is then likely to lead to the value of $\bar{\tau}_{\rm ff}$ at 15 GHz being much lower in Cyg X-3 than in Cyg X-1, in spite of the wind mass loss rate being about an order of magnitude higher. 

Given that, we expect the value of $r$ in Cyg X-3 to be much closer to unity than that in Cyg X-1. Thus, the above effect of the intrinsic correlation index being lower than the observed one is probably minor in Cyg X-3. Therefore, it is unlikely that the value of $p$, which we have estimated correcting for the X-ray absorption, is significantly higher than our value of $\simeq$1.5. 

\subsection{The nature of the hard state and comparison to other X-ray binaries}
\label{xray}

The anticorrelation between hard and soft X-rays within the hard state of Cyg X-3 is a consequence of the soft spectrum becoming softer when brighter. This leads to a pivoting in the $\simeq$10--20 keV energy range. Then, a flux increase in soft X-rays corresponds to a decrease in hard X-rays, and vice versa. This pivoting pattern is seen, e.g., in fig.\ 2 of SZM08, see their spectra 1 and 2. 

On the other hand, Cyg X-1 has only moderate spectral variability in the hard state. Based on the energy dependence of the fractional variability, see, e.g., fig.\ 12 in \citet{z02}, the hard-state pivot is at $\sim$100 keV. Consequently, there is no observed anticorrelation between the soft and hard X-rays in the hard state, as found by ZSPL11. 

As calculated in \citet{z02}, the X-ray variability patterns in the hard state are well explained by originating from thermal Comptonization. If the dominant driver of the variability is in a changing flux of seed photons, the spectrum softens with the increase of the total flux, leading to a pivoting behaviour. The lower value of the pivot in Cyg X-3 than in Cyg X-1 is most likely related to the hard state in the former being in fact intermediate. In Cyg X-1, the transition from the hard to the intermediate state is indeed characterized by pivoting around $\sim$10--20 keV, see fig.\ 13 in \citet{z02}. 

Given our results in Section \ref{rx} and the discussion in Section \ref{intrinsic}, the absorption in Cyg X-3 has to be strong, and the intrinsic soft X-ray spectrum in the hard state has to be rather soft, $\Gamma\sim 2$. Thus, the hard state of Cyg X-3 corresponds indeed to the intermediate state of BH binaries. As found by ZSPL11 (see their figs.\ 8a, b), the positive correlation between the radio and bolometric X-ray fluxes in Cyg X-1 is present in both the hard and intermediate states, and breaks down only in the soft state. We thus expect a similar situation in Cyg X-3, which can explains the observed correlation even if its spectral state is intermediate.   

If the apparent hard state in Cyg X-3 is intrinsically intermediate, this can explain the difference in its spectral shape with respect to BH binaries in the hard state without scattering in the stellar wind, invoked by \citet{zmg10}. The apparent peak at $\sim$10 keV in the hard-state $E F(E)$ spectrum is then an artefact of strong absorption, and the spectrum at higher energies is due to non-thermal, rather than thermal, Comptonization \citep{h08}. Also, the hard state of Cyg X-3 appears then to be similar to the hardest spectral state of the BH LMXB GRS 1915+105, see, e.g., fig.\ 4 in \citet*{dwg04}, where the hardest spectra are similar to the intrinsic hard-state spectra of Cyg X-3 shown in Figs.\ \ref{spectra}(b,d).

In order to determine the relationship of Cyg X-3 to other accreting X-ray binaries, we compare quantitatively their overall X-ray and radio properties. We first compare the positions on the X-ray colour-colour and colour-luminosity diagrams. We use the compilations of the data from \citet{dg03}, \citet*{dgk07}, \citet{dwg04} and \citet{gdg07} for Galactic BH and weakly-magnetized NS binaries. Those authors used the soft and hard colours defined by the energy flux ratio of the bands 4--6.4 keV to 3--4 keV and 9.7--16 keV to 6.4--9.7 keV, and the bolometric luminosity in the Eddington units. We have updated the values of the distances and BH masses according to the current best estimates, in particular those of \citet{cj14} and \citet{cs15}. We calculate the colours and Eddington ratios, $L_{\rm bol}/L_{\rm Edd}$, for Cyg X-3 assuming $d=8$ kpc, $M=4\msun$ \citep{zmb13} and using the ASM-BAT absorption-corrected spectra shown in Fig.\ \ref{spectra}(b). Here $L_{\rm Edd}$ is the Eddington luminosity, calculated for pure He ($X=0$) for Cyg X-3, and for the cosmic composition ($X=0.7$) for other binaries. To calculate the fluxes within a part of a bin, we assume the spectral shape within a bin to be flat in $E F(E)$. The spectra at $E< 1.3$ keV are most likely dominated by disc blackbody emission, and we thus add a contribution from those energies assuming $F(E)\propto E^{1/3}$. We obtain the bolometric fluxes of 2.3, 5.5 and $8.0\times 10^{-8}$ erg cm$^{-2}$ s$^{-1}$ in the hard, intermediate and soft state, respectively, which correspond to the bolometric (isotropic) luminosities of 1.8, 4.2, $6.2\times 10^{38}(D/8\,{\rm kpc})^2$ erg s$^{-1}$, and the Eddington ratios of 0.18, 0.42 and $0.62 (D/8\,{\rm kpc})^2 (M/4\msun)^{-1}$, respectively. We can see in Fig.\ \ref{lcd} that indeed the colours and Eddington ratios of the spectral states of Cyg X-3 (at $4\msun$) are within the range occupied by GRS 1915+105 (using the data of \citealt{dwg04}), on both the colour-colour and colour-$L_{\rm bol}/L_{\rm Edd}$ diagrams.

On the other hand, \citet{h09} used similar diagrams and argued that the most likely BH mass is $30\msun$. In fact, that mass may also be compatible with our values of colour-$L_{\rm bol}/L_{\rm Edd}$, and it would correspond to moving the points on Fig.\ \ref{lcd}(b) down to a region populated by other BH binaries, but not GRS 1915+105. However, the Cyg X-3 points on the colour-colour diagram would still be within those of GRS 1915+105, which represents an inconsistency. Also, such a high BH mass is not found in other BH binaries (see \citealt*{lmc15} for the lack of dynamical evidence for a high mass BH in IC10 X-1), as well as it is in contradiction with the later results of \citet{zmb13}. Compared to \citet{h09}, the absorber in Cyg X-3 adopted by us is stronger, which would move the values of the colours in that paper to lower values, making them more compatible with those of GRS 1915+105. We also note that the distance to GRS 1915+105 has been revised down from 12.5 kpc to $8.6^{+2.0}_{-1.6}$ kpc \citep{reid14}, which has decreased its Eddington ratios by a factor of 2.1 (at the best-fit distance). This results in our Cyg X-3 points for $M=4\msun$ to be entirely within the region occupied by GRS 1915+105.

We have also compared the colours and the Eddington ratios of Cyg X-3 to those of weakly magnetized NS binaries, assuming now $M=1.5\msun$. We have found the results of this comparison compatible with Cyg X-3 containing an NS, with its colours and Eddington ratios similar to those of Z sources. However, the strength of the radio emission relative to the X-ray in Cyg X-3 exceeds that of all of BH binaries, see below, and it is much much higher than for NS ones \citep{mf06,corbel13a}.

\begin{figure}
\centerline{\includegraphics[width=\columnwidth]{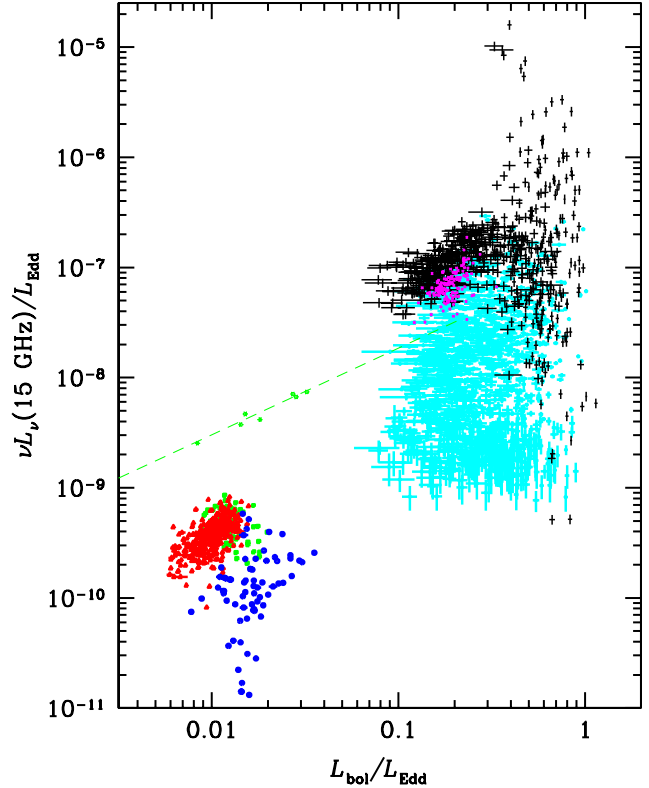}
}
\caption{The correlation between the unabsorbed bolometric luminosity in Eddington units with the radio 15 GHz luminosity for Cyg X-3 (8 kpc, $4\msun$; black error bars) compared to that of Cyg X-1 (1.86 kpc, $16\msun$; the red triangles, green squares and blue circles for the hard, intermediate and soft state, respectively, from ZSPL11), GX 339--4 (8 kpc, $12\msun$; green squares fitted by the green dashed line, from \citealt{z04}, but multiplied by 1.25 to account for those measurements being at 8.6 GHz) and GRS 1915+105 (8.6 kpc, $12.4\msun$; cyan squares with error bars), with the magenta squares corresponding to the plateau $\chi$ state using the criteria described in Section \ref{xray}.
}
\label{rbol}
\end{figure}

We then compare the radio vs.\ the bolometric luminosity relationship of Cyg X-3 with those of selected BH binaries, see Fig.\ \ref{rbol}, where the $y$ axis shows $\nu L_\nu/L_{\rm Edd}$ at 15 GHz. The black error bars shows the points for Cyg X-3 at $M=4\msun$ and $d=8$ kpc, where the shown values of $L_{\rm bol}$ have been corrected with respect to $L(1.3$--$100\,{\rm keV})$ in the same way as described above for Fig.\ \ref{lcd}(b). 

We compare first Cyg X-3 with GX 339--4, for which the radio/X-ray correlation is the best studied one \citep{corbel13a}. Unfortunately, the published correlations are given for narrow bands of either 3--9 keV or 1--10 keV only. As we have argued above, a narrow-band X-ray flux is likely to be significantly affected by spectral variability (as it is very significantly the case in Cyg X-3, and significantly in Cyg X-1, see ZSPL11). The only radio/$L_{\rm bol}$ correlation available for an LMXB appears to be that of \citet{z04} for GX 339--4, which we show in green points. The magenta line shows the best fit power-law dependence, with $p\simeq 0.79\pm 0.07$. The radio fluxes are at 8.6 GHz \citep{corbel03}, which we multiply by a factor of 1.25 in order to estimate the 15 GHz flux using the average $\alpha\simeq 0.4$ observed in the hard state of GX 339--4 \citep{corbel13b}. On the other hand, we have checked that the 8.3--15 GHz index of Cyg X-3 is $\simeq 0$ in its hard state, cf.\ Figs.\ \ref{alpha}(a--b), and it is $\simeq 0$ for Cyg X-1 in the hard state \citep{fender00}. Thus, the effect of the radio band on the presented correlation appears minor. We see Cyg X-3 in its hard state is significantly more radio-loud than GX 339--4, by a factor of $\simeq$4 on average (though there is a systematic uncertainty concerning the correction for the intrinsic X-ray absorption in Cyg X-3). 

As we have shown above, Cyg X-3 appears relatively similar spectrally in X-rays to GRS 1915+105. In order to them in the radio band, we use the intense Ryle telescope monitoring of GRS 1915+105 conducted during MJD 49856--53898, and the ASM data, during MJD 50088--55859. Given the fast variability of that object, we have tested the effect of various averaging methods. We have finally chosen to average the radio over observation periods containing gaps shorter than 0.2 d, which resulted in the average length of an averaged segment of 0.09 d (usually one per day) with $\sigma\simeq 0.07$ d. We then correlated those data with the ASM dwell data which were within $\pm 0.2$ d from the average time of a given radio segment. We have corrected the averaged ASM data for line-of-sight absorption assuming $N_{\rm H}=3.8\times 10^{22}$ cm$^{-2}$ \citep{ebisawa98,cc04} and the intrinsic spectra with $\Gamma=2$ within each ASM bin. We have further added the contribution from $E<1.3$ keV in the same way as for Cyg X-3, i.e., assuming $F(E)\propto E^{1/3}$, see above. We have also considered the available BATSE data for this object\footnote{\url{http://gammaray.nsstc.nasa.gov/batse/occultation/}}. We have found that addition of the 20--160 keV fluxes to those of the ASM was possible only for about 40 per cent of the ASM points, and an upper part of the correlation was not covered by the ASM+BATSE points. The average 20--160 keV to 1.3--12.2 keV flux ratio equals 0.11. Given that the $EF(E)$ spectra of GRS 1915+105 in almost all states are decreasing above several keV \citep{dwg04}, we estimate the flux contribution from 12--20 keV to be similar to that from $>20$ keV. We have thus applied an approximate bolometric correction of a factor of 1.2 to the fluxes from the ASM. 

We show the results by the cyan error bars in Fig.\ \ref{rbol}. We see GRS 1915+105 does not show a distinct radio/X-ray correlation. On the other hand, \citet{rushton10} claimed a strong correlation within the plateau (steady and hard) state of GRS 1915+105. When using the same time intervals as those given in \citet{rushton10} as used for their correlation, we have found a rather scattered diagram, with a steep positive slope at high radio fluxes and a negative one at low radio fluxes. When we used the same criteria as given in that paper, i.e., the ratio of the count rates in the ASM third to second channels of $>1.5$ and the total ASM rate of 30--50 s$^{-1}$, we found a similar discrepancy. We have recovered a correlation only after imposing an additional criterion of the radio flux of $>40$ mJy (which, however, implies that the X-ray state by itself does not set the radio behaviour, in contrast to Cyg X-3). Those points are shown in Fig.\ \ref{rbol} by green crosses. We do see some correlation, with the slope of $2.1\pm 0.2$, somewhat steeper than $1.7\pm 0.3$ given in \citet{rushton10}. 

We see that GRS 1915+105 occupies a similar range of $L_{\rm bol}/L_{\rm Edd}$ as Cyg X-3 (as also seen in Fig.\ \ref{lcd}b), but it is both less radio loud and with a much less clear radio/X-ray correlation. The likely reason for that is the difference in the jet environments. Cyg X-3 is a binary with a high-mass donor, and interaction between its wind and the jet leads to presence of shocks, in particular of recollimation shocks \citep*{yh15,zyh15}. Those shocks are likely to efficiently accelerate electrons, leading to enhanced radio emission. However, no such interaction takes place in GRS 1915+105, an LXMB.

On the other hand, Cyg X-1, which is also a high-mass X-ray binary, has the hard-state radio fluxes, shown by the red triangles in Fig.\ \ref{rbol}, an order of magnitude below those of the GX 339--4 correlation. An explanation for that may be provided by the wind ram pressure around the jet in Cyg X-1 being much lower than that in Cyg X-3, owing to both the lower mass-loss rate and the larger separation. The jet lateral pressure in Cyg X-1 may then be always larger than that of the wind, provided the jet power is above some critical value \citep{zyh15}. Then no recollimation shock would form. 
 
Furthermore, as discussed in Section \ref{intrinsic}, the radio flux in Cyg X-1 is significantly free-free absorbed in the wind from its donor, see ZSPL11 and \citet{zdz12}. In the model of the latter, the slope of the intrinsic hard-state correlation in Cyg X-1 is close to that of the lower branch of the radio/X-ray correlation, first found for H1743--322 \citep{coriat11}. However, the correction for the wind absorption in Cyg X-1 is relatively uncertain. More detailed knowledge of the wind structure in Cyg X-1 is needed to determine the degree of free-free absorption.

\subsection{Consequences for radio/X-ray correlations in other objects}
\label{other}

Our results point out to the importance of taking into account an estimate of the bolometric flux when correlating X-rays with radio, reinforcing the similar conclusion of ZSPL11. Due to the pivoting spectral variability in X-rays the hard X-rays are often a very bad predictor of the bolometric flux, and can be either uncorrelated or negatively correlated with it. This effect appears to be present also in Seyfert galaxies, in which a common variability pattern in soft X-rays is the steepening of their soft X-ray spectrum with the increasing flux. This then leads to a pivot in hard X-rays. 

This effect can explain the discrepancy between the result of \citet*{salvato04}, who found the radio and soft X-ray emission of their sample of Seyferts to be strongly correlated, and the result of \citet{burlon13}, who found a lack of such correlation for hard X-rays, using \swift/BAT data. The latter authors doubted then the reality of the soft X-ray correlation. However, this discrepancy can be resolved by a pivoting spectral variability of X-rays. In Cyg X-3, it leads to the 10--20 keV flux to be completely uncorrelated with radio, see Fig.\ \ref{rx4}(g), in spite of the correlation in soft X-rays, Fig.\ \ref{rx4}(e), being very strong. 

\section{Conclusions}
\label{conclusions}

We have studied implications of the presence of the positive and negative correlation with the radio flux for the soft and hard X-ray fluxes, respectively, in the hard spectral state of Cyg X-3. We have found that the radio flux is then completely uncorrelated with the observed, absorbed, broad-band, 1.3--100 keV X-ray flux, which appears to be a good approximation to the absorbed bolometric flux. An intrinsic lack of correlation is unlikely on theoretical grounds, as well as it is unlikely given the presence of the strong narrow X-ray band correlations. 

However, we can recover a clear positive correlation if the absorption by the stellar wind of the donor in Cyg X-3 is strong enough. This implies that the intrinsic soft X-ray spectrum in the hard state is relatively soft, similar to those of intermediate states of BH binaries. This nature of the apparent hard state of Cyg X-3 is also consistent with the luminosity of the hard state being much higher in this object than in the hard state of the high-mass BH binary, Cyg X-1. 

We have found the radio spectra in the hard state to be hard on average, with $\alpha\ga 0$. Such spectra are characteristic to BH X-ray binaries in the hard state.

We have compared Cyg X-3 with other X-ray binaries. On the X-ray colour-colour and colour-Eddington ratio diagrams, Cyg X-3, after correcting its spectra for absorption and for the BH mass of $\sim 4\msun$, appears similar to GRS 1915+105, with similar colours and $L_{\rm bol}/L_{\rm Edd}\sim 0.1$--1. On the other hand, Cyg X-3, at $1.5\msun$ is also similar on such diagrams to NS Z-type binaries. However, the radio emission of those sources is much weaker than that of Cyg X-3, which is both the most radio-luminous and, in its hard state, the most radio-loud (i.e., with the highest ratio of the radio to bolometric luminosities) source among all known X-ray binaries. Still, its radio loudness is much closer to those of BH binaries than to those of NS ones. This represents a strong argument for the presence of a BH in Cyg X-3. We interpret the radio loudness and radio fluxes of Cyg X-3 being much higher than those of GRS 1915+105 as due to an enhancement of the jet emission in the former due to interaction of the jet with the very strong stellar wind from the donor. 

We have found (Appendix \ref{lc}) that the flux distributions of the radio emission and soft X-rays can be described by two log-normal functions, one corresponding to the hard state and one to soft state. This indicates a multiplicative and correlated character of variability of Cyg X-3 in both the hard and soft state and in both the accretion flow and the jet. 

Finally, we have presented (Appendix \ref{lc}) the long-time light curves of Cyg X-3 using the available data, and compared them to the occurrences of high-energy \g-ray detection. We have found that a low hard X-ray flux, $F/\langle F\rangle\la 0.1$, is both a necessary and sufficient criterion for detectable \g-rays. We have found that this criterion has not been satisfied since MJD 55650 (2011 March 30) until now, MJD 57333.

\section*{ACKNOWLEDGMENTS}

We thank Patryk Pjanka for help with the data analysis, Antonino D'Ai for valuable comments, and Chris Done and Marek Gierli{\'n}ski for providing us with the colour-flux data on BH binaries. This research has been supported in part by the Polish NCN grants 2012/04/M/ST9/00780 and 2013/10/M/ST9/00729. The Ryle Telescope and the AMI arrays are supported by STFC and the University of Cambridge. This research has also made use of the MAXI data provided by RIKEN, JAXA and the MAXI team, and of the quick-look results provided by the \xte\/ ASM team.

\appendix

\section{The flux distributions and light curves of Cyg X-3}
\label{lc}

\begin{figure*} 
\centerline{\includegraphics[width=7.cm]{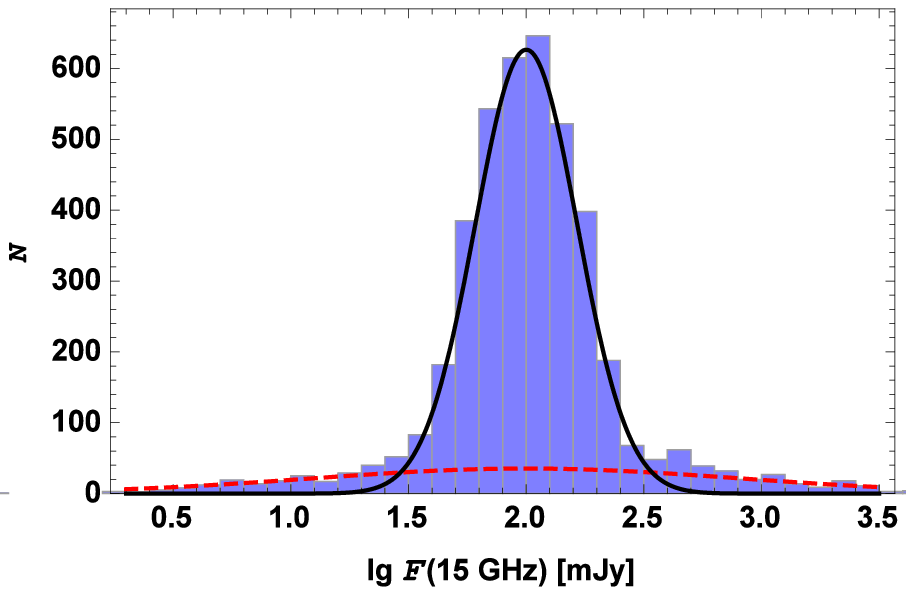}
\includegraphics[width=7.cm]{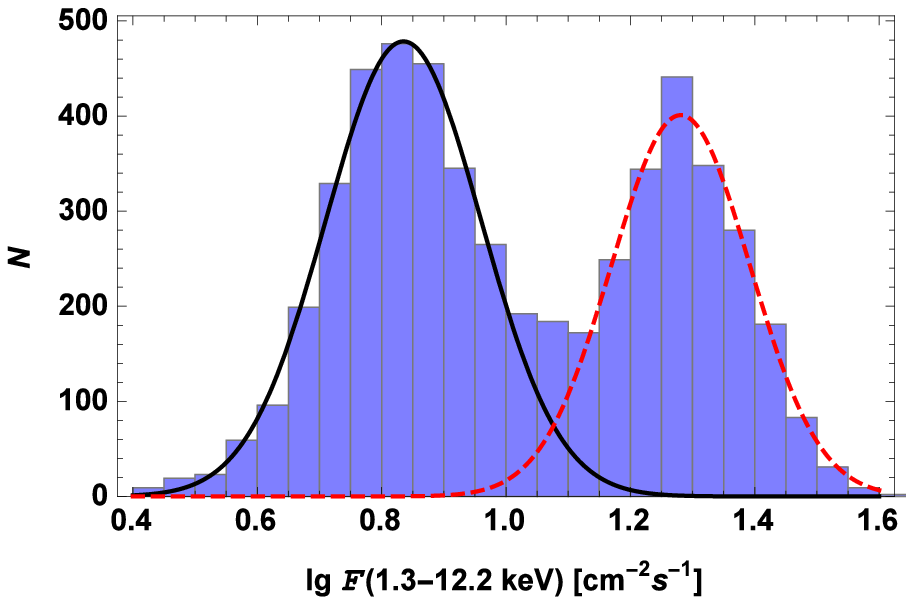}}
\centerline{\includegraphics[width=7.cm]{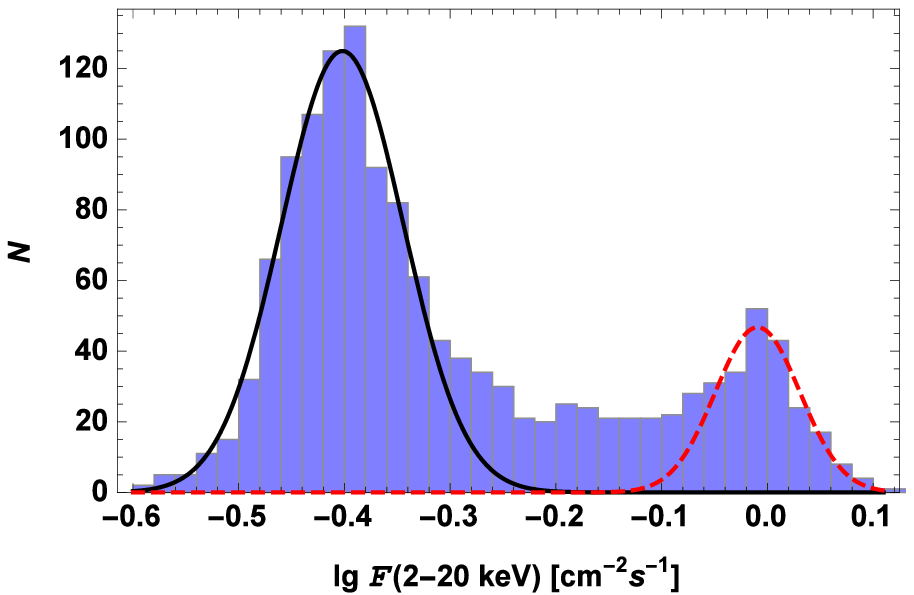}
\includegraphics[width=7.cm]{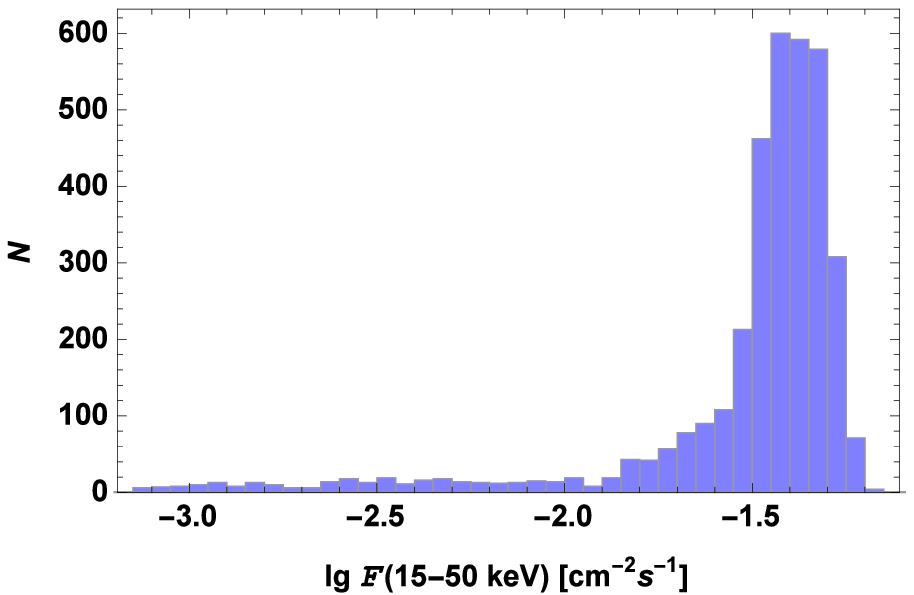}}
\caption{The histograms of the 15 GHz fluxes and the ASM, MAXI and BAT count rates, from top left to bottom right. The first three histograms are well represented by a sum of two log-normal distributions, shown by the black solid and red dashed curves, which correspond to the hard and soft state, respectively.
} 
\label{hist}
\end{figure*}

We show here distributions of the fluxes and the light curves for the X-ray monitors used here and for the Ryle/AMI. Fig.\ \ref{hist} shows the histograms of the 15 GHz, ASM, MAXI and BAT fluxes and count rates (for simplicity, we use the symbol $F$ for both). The histogram for the BATSE flux is shown in \citet{h08}. The rates for 15 GHz, and ASM and MAXI are fair estimators of the jet power in the former case and the accretion power in the latter, see Section \ref{rx}. On the other hand, the hard X-rays are anticorrelated with the accretion luminosity in the hard/intermediate state due to the spectral pivoting. Thus, their distribution does not reflect the accretion power. 

Consequently, we find the 15 GHz, ASM and MAXI histograms can be fitted by a sum of two log-normal distributions, one present in the hard/intermediate state and one in the soft state, while it is not the case for the BAT histogram. The values of the mean and standard deviation for the hard/intermediate-state and soft-state log-normal distributions for the 15 GHz, ASM and MAXI data are $(-2.00,0.22;\, -2.09,0.90)$, $(0.84,0.13;\, 1.28,0.11)$, $(-0.40,0.06;\, -0.01,0.04)$, respectively (in $\log_{10}$ units). We see that the width of the flux distribution in each of the state in the ASM data is twice as large as that of the MAXI data. This is due to the MAXI 10-20 keV channel containing the pivot of the spectral variability, see Fig.\ \ref{rx4}(g), which reduces the range of the flux variability. On the other hand, the ASM 5--12.2 keV channel is still below the pivot, see Fig.\ \ref{rx4}(c). All of the ASM and MAXI distributions are narrower than the corresponding 15-GHz ones, especially in the soft state. This reflects the dramatic variability of the radio emission observed in that state.

The presence of a log-normal distribution shows that the underlying physical process has to be multiplicative, with variations coupled together on all time scales \citep*{uttley05}. Interestingly, we find the log-normal variability to be present in the both main states of Cyg X-3, hard and soft. On the other hand, \citet{uttley05} and \citet*{pzi08} have found it in the hard state of Cyg X-1 on $\sim$0.1--10 s (for soft X-rays) and $\ga 0.1$ d (for soft X-rays and radio) time scales, respectively. However, it appears that a log-normal flux distribution has not been found before in the soft state of an X-ray binary.

Fig.\ \ref{lc901} shows the light curves. Since the variability in all energy bands spans a large ratio and it is log-normal for some detectors, we normalize each light curve to its geometric unweighted average over the respective observation time span, i.e., we calculate $\langle \ln F\rangle$. For that, we have to select only points with the measured fluxes of $F_i>0$. This will lead to a certain overestimate of the overall average, but the effect is minor for the light curves presented below. For notational simplicity, we denote $\langle F\rangle \equiv \exp\, \langle \ln F\rangle$. Also, in order to reduce a contribution from the measurements of poor quality, we include only data with the fractional error of $\Delta F_i/F_i<0.5$, where $\Delta F_i$ is the flux error. We show the 1-d average rates except for the BATSE, where we show the 3-d average. In the case of BAT, we use the public 15--50 keV data\footnote{\url{http://swift.gsfc.nasa.gov/results/transients/CygX-3/}}  \citep{krimm13}. That light curve is very similar to our 14--50 keV one. For the MAXI, we show the 2--10 keV light curve instead of the full 2--20 keV one in order to have as much correspondence as possible to the ASM 1.3--12.2 keV light curve. 

The average rates and standard deviations are given in Table A1. The standard deviations correspond to the excess variance, i.e., the variance due to the measurement errors is subtracted,
\begin{equation}
\sigma^2={1\over N}\sum_i \left[\ln^2 F_i -\langle \ln F\rangle^2 -(\Delta F_i/F_i)^2 \right].
\label{sigma}
\end{equation}
The lower value of $\sigma$ for the BATSE data than that for the BAT is probably due to the former data being averaged over 3-d rather than 1-d intervals. The large values of $\sigma$ for all the radio data is the consequence of the dramatic radio variability in this object, showing both very bright flares and very weak states, see Fig.\ \ref{lc901}. For the 15 GHz data, we have used the estimate of $\Delta F_i/F_i= 0.1$ (Section \ref{data}). For clarity, we do not show the GBI light curves, which are relatively similar to the Ryle/AMI one. 

We also mark the intervals corresponding to the detected high-energy \g-ray emission in Fig.\ \ref{lc901}. These are MJD 54566--54647, 54821--54850, 55324--55326 \citep{bulgarelli12}, 54750--54820, 54990--55045 \citep{fermi09}, 55343--55345 \citep{williams11,bulgarelli12}, 55586--55610, 55642-55644 \citep{corbel12} and MJD 57398--57412 \citep{fermi16,agile16}. All those periods correspond to the soft state as defined in our paper, and, with exception for the last occurence, to periods with both the lowest 15--50 keV flux detected, with $F/\langle F\rangle \la 0.1$, and occurrences of radio flares, with the 1-d averaged 15 GHz flux of $F/\langle F\rangle \ga 5$ (as noted, e.g., by \citealt{corbel12}; \citealt{bulgarelli12}). However, the former of these conditions, i.e., a low hard X-ray flux, appears to be both necessary and sufficient. We note that this condition has not been satisfied during MJD 55650--57397, see Fig.\ \ref{lc901}. Indeed, no high-energy \g-ray detection from Cyg X-3 has been reported during that period.

\setlength{\tabcolsep}{3.5pt}
\begin{table}
\begin{center}
\caption{The logarithmically-averaged detector rates and standard deviations for Cyg X-3.}
\begin{tabular}{ccccc}
\hline
Detector & Energy range & Unit & $\langle F\rangle$ & $\sigma$ \\
\hline
GBI            & 2.25 GHz    & mJy                         & 99     & 0.91\\
GBI            & 8.3 GHz     & mJy                         & 125    & 0.84\\
Ryle, AMI      & 15 GHz      & mJy                         & 103    & 0.94\\
\xte/ASM       & 1.3--12.2 keV & count\,s$^{-1}$             & 10.5   & 0.58\\
{\it ISS}/MAXI & 2--10 keV   & photon\,cm$^{-2}$\,s$^{-1}$ & 0.350  & 0.50\\
\swift/BAT     & 15--50 keV  & count\,cm$^{-2}$\,s$^{-1}$           & 0.0336 & 0.54\\
\gro/BATSE     & 20--100 keV & keV\,cm$^{-2}$\,s$^{-1}$    & 1.38   & 0.30\\
\hline
\end{tabular}
\end{center}
\label{averages}
\end{table}

\begin{figure*}
\centerline{\includegraphics[height=7.3cm]{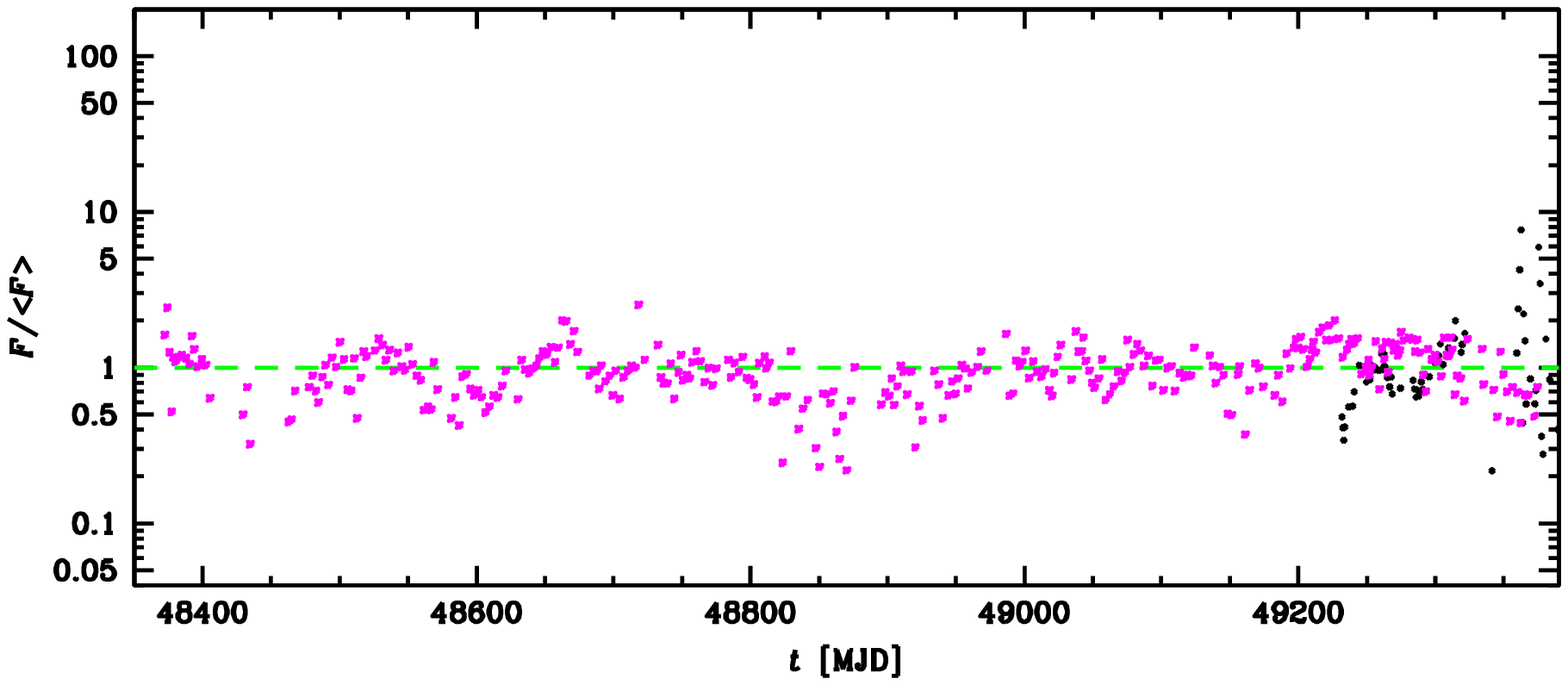}}
\centerline{\includegraphics[height=7.3cm]{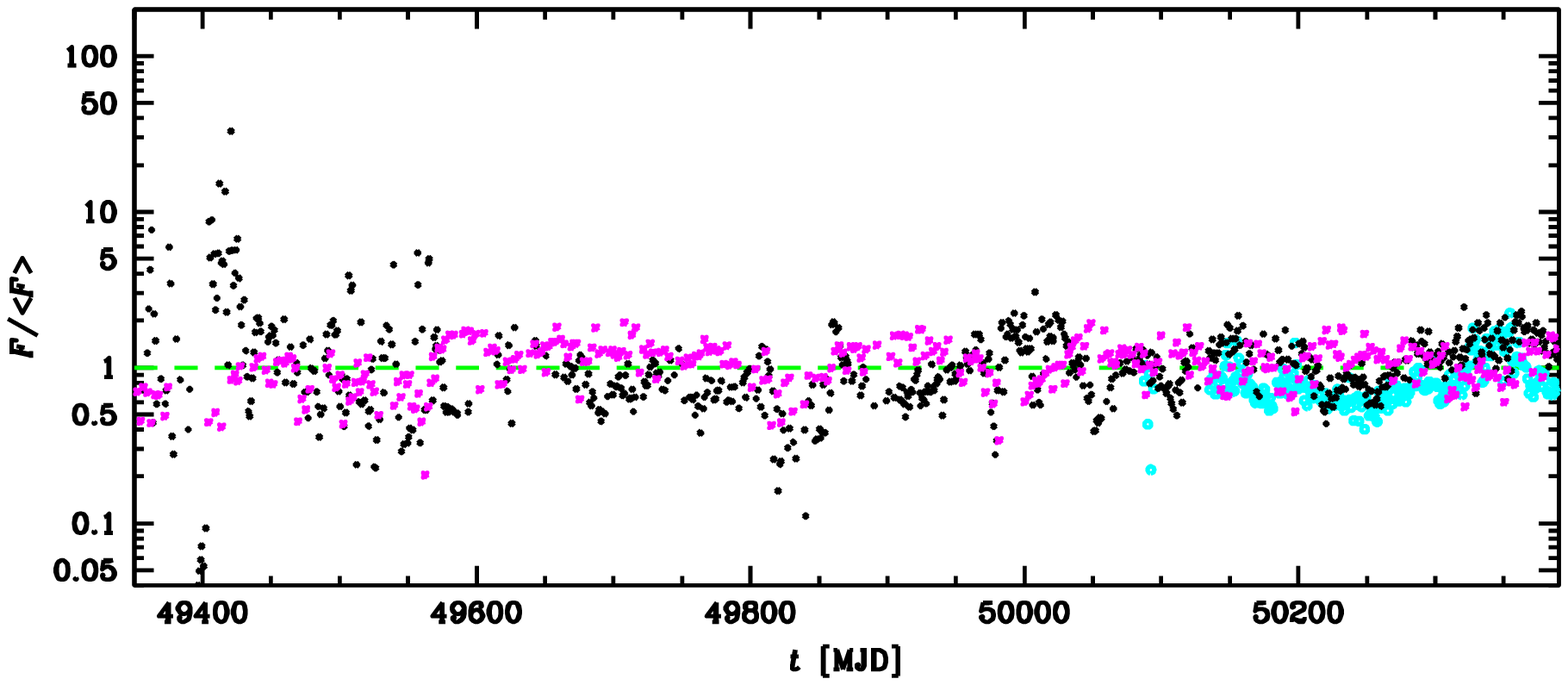}}
\centerline{\includegraphics[height=7.3cm]{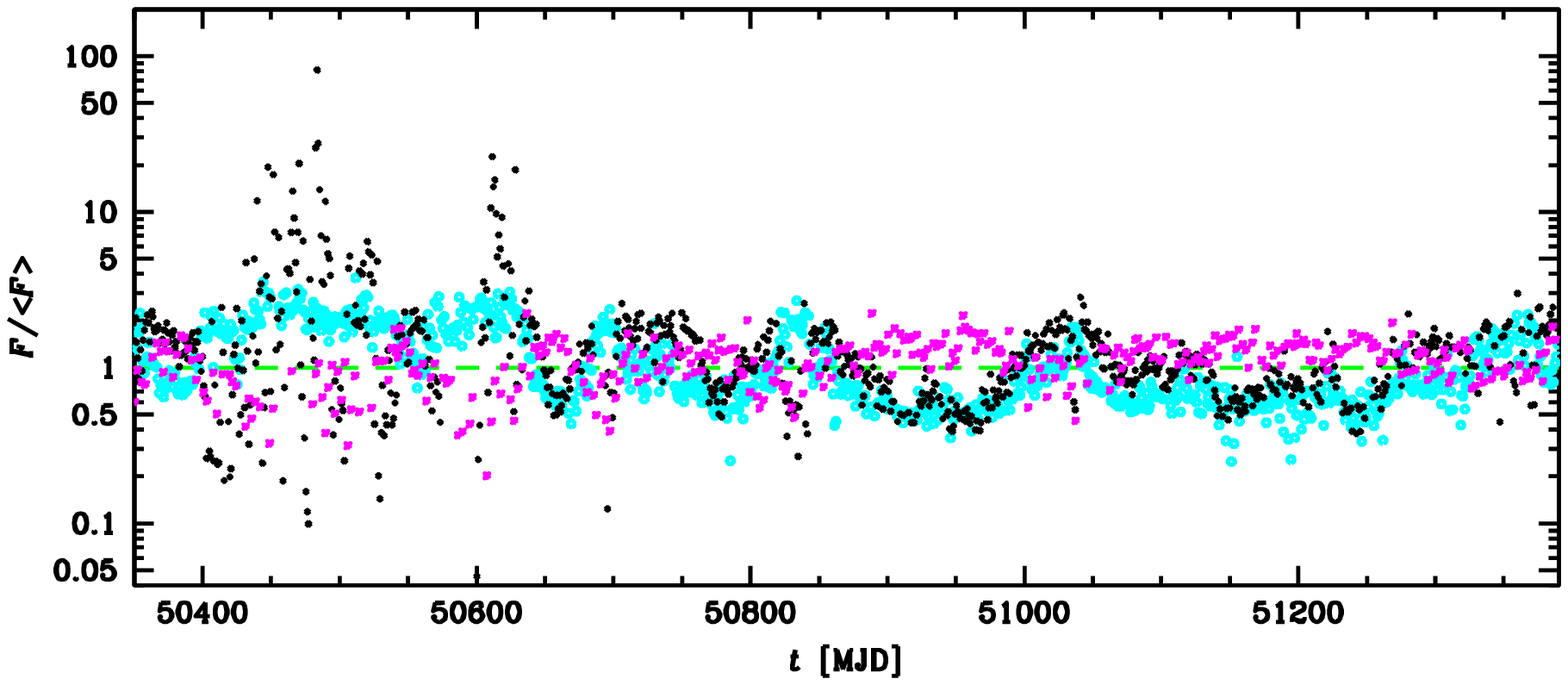}}
\caption{The long-term light curves of Cyg X-3 normalized to the respective average over the observation length. The cyan, blue, red, magenta and black symbols show the rates normalized to the respective average for the ASM (1.3--12.2 keV), MAXI (2--10 keV), BAT (15--50 keV), BATSE (20--100 keV) and Ryle/AMI (15 GHz). For clarity of display, the error bars are not plotted. The dashed green line corresponds to the average. The heavy magenta horizontal lines correspond to the detected 8 occurrences of high-energy \g-ray emission.
}
\label{lc901}
\end{figure*}

\setcounter{figure}{1}
\begin{figure*}
\centerline{\includegraphics[height=7.3cm]{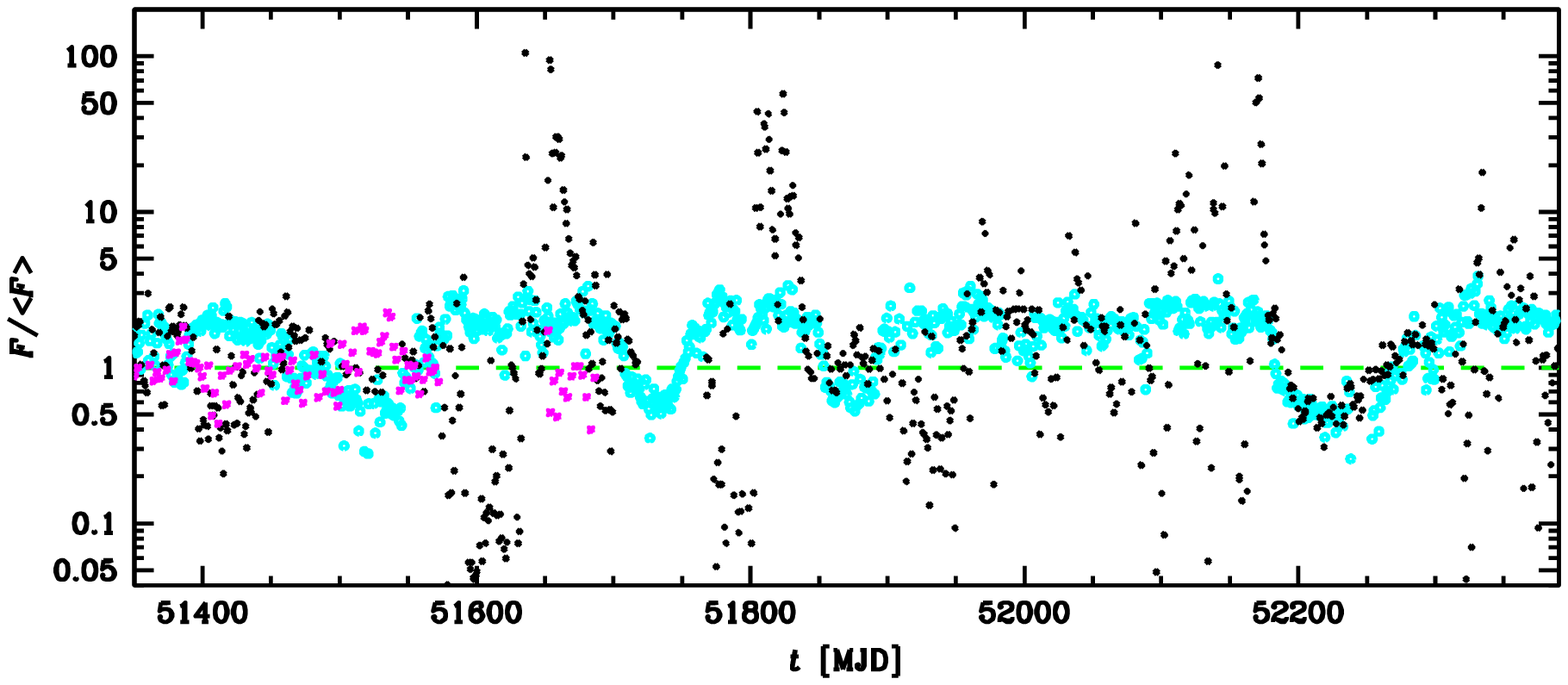}}
\centerline{\includegraphics[height=7.3cm]{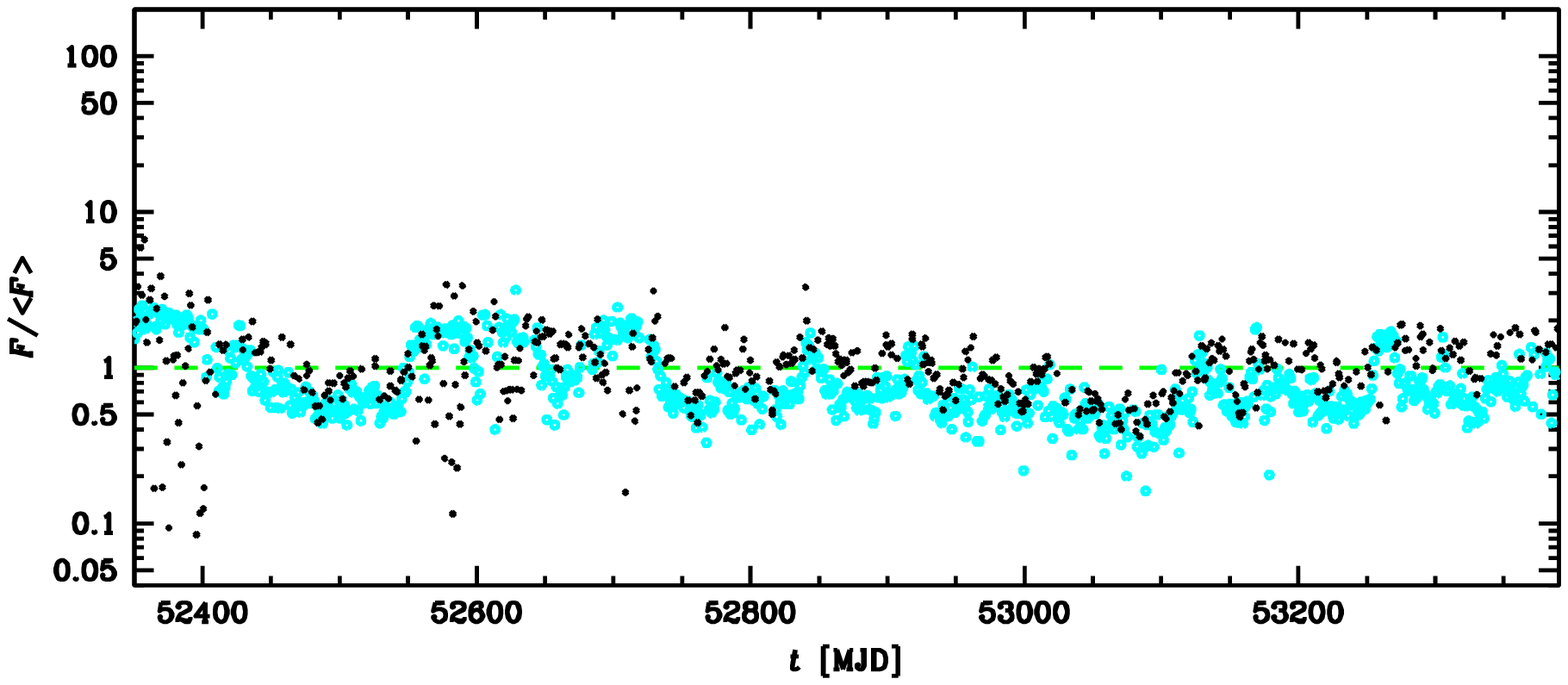}}
\centerline{\includegraphics[height=7.3cm]{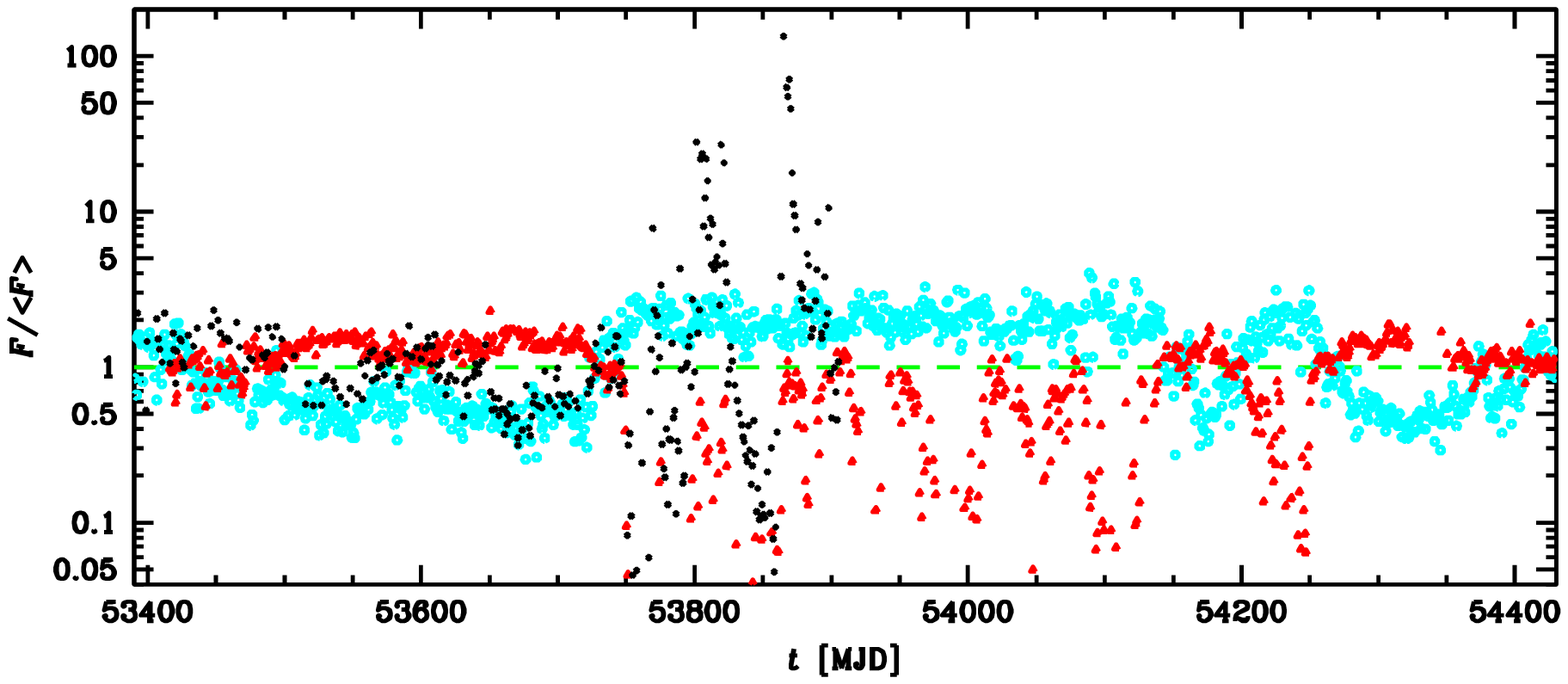}}
\caption{Continued. 
}
\label{lc234}
\end{figure*}

\setcounter{figure}{1}
\begin{figure*}
\centerline{\includegraphics[height=7.3cm]{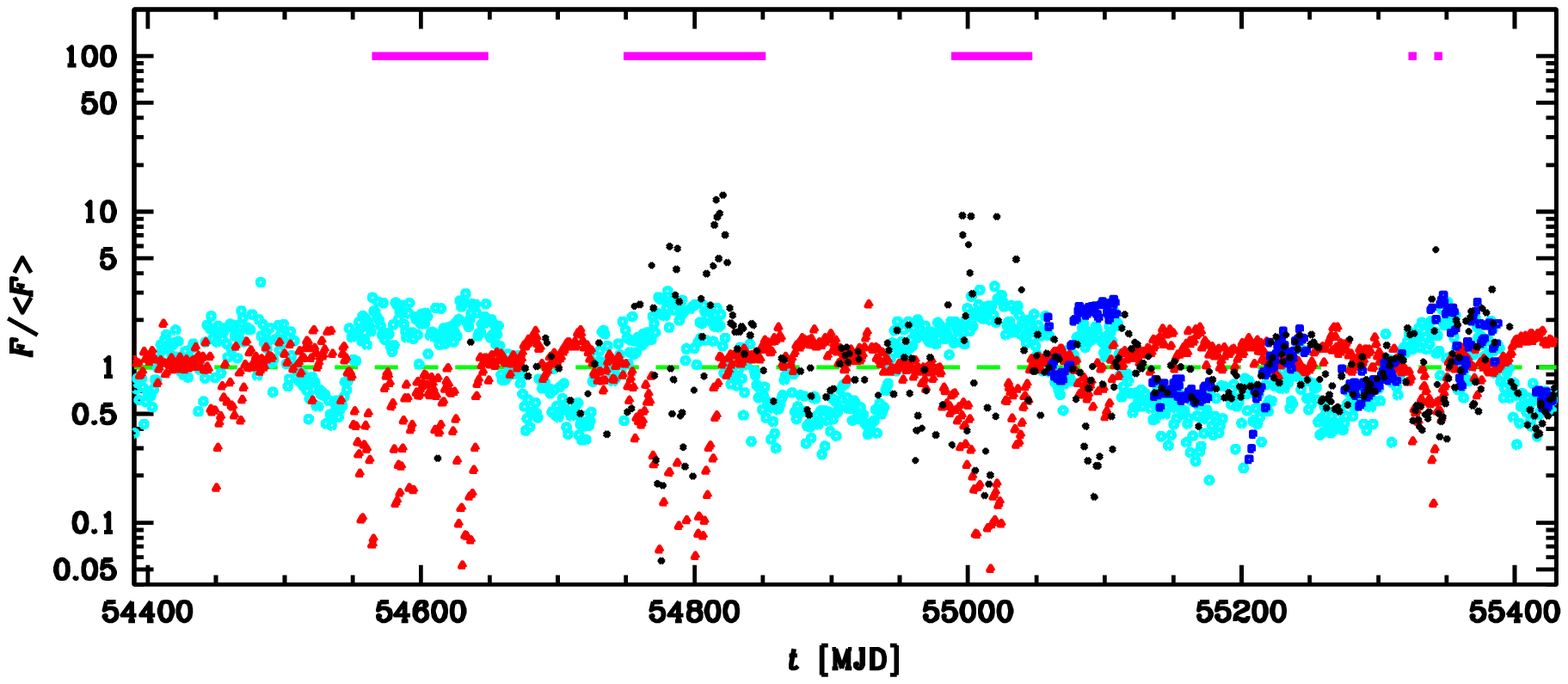}}
\centerline{\includegraphics[height=7.3cm]{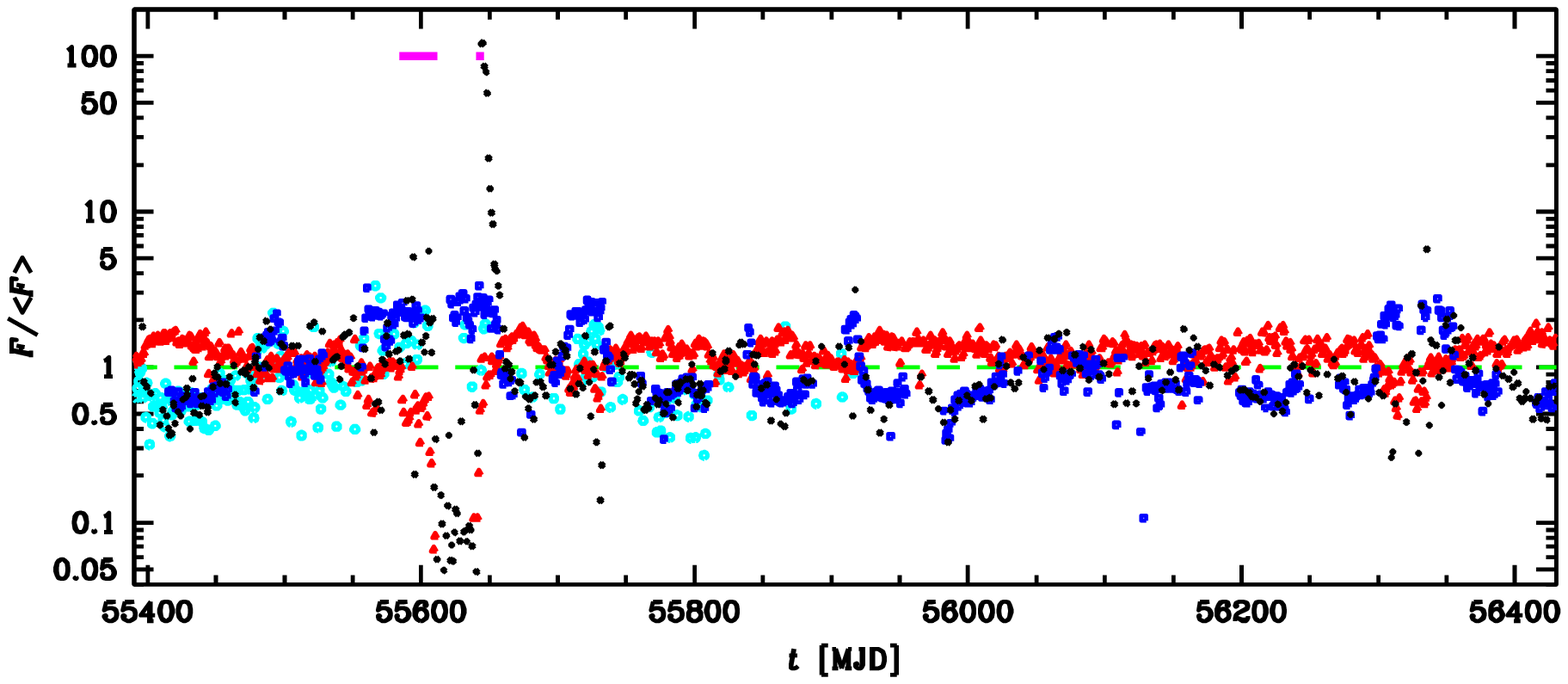}}
\centerline{\includegraphics[height=7.3cm]{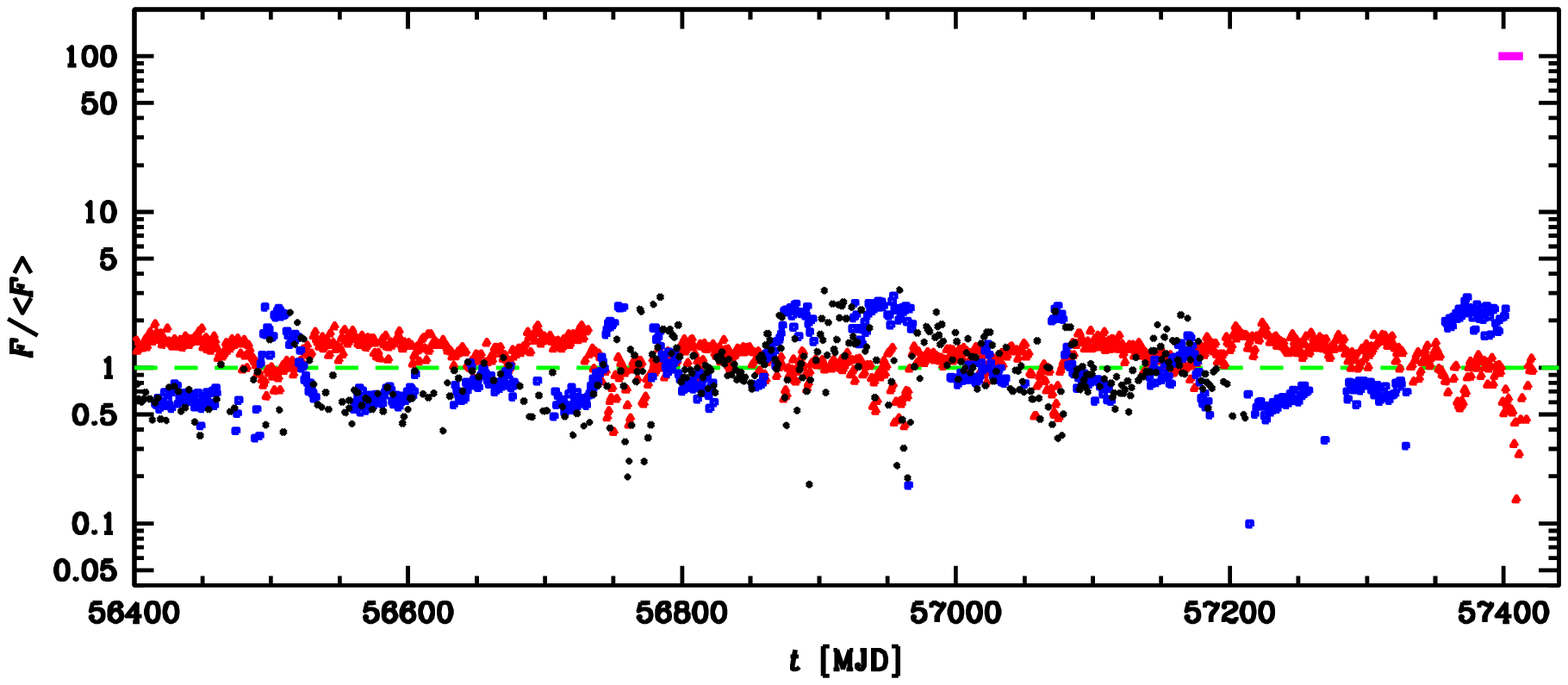}}
\caption{Continued. 
}
\label{lc56}
\end{figure*}

\label{lastpage}

\begin{thebibliography}{}

\bibitem[\protect\citeauthoryear{Arnaud}{1996}]{arnaud96} 
Arnaud K. A., 1996, in Jacoby G. H., Barnes J., eds., Astronomical Data Analysis
Software and Systems V, ASP Conf. Series Vol.\ 101, San Francisco, p.\ 17

\bibitem[\protect\citeauthoryear{Axelsson, Larsson \& Hjalmarsdotter}{Axelsson et al.}{2009}]{alh09} 
Axelsson M., Larsson S., Hjalmarsdotter L., 2009, MNRAS, 394, 1544

\bibitem[\protect\citeauthoryear{Barthelmy et al.}{2005}]{barthelmy05}
Barthelmy S. D. et al., 2005, Space Sci. Rev., 120, 143

\bibitem[\protect\citeauthoryear{Bautista \& Kallman}{2001}]{bk01} 
Bautista M.~A., Kallman T.~R., 2001, ApJS, 134, 139 

\bibitem[\protect\citeauthoryear{Belczy{\'n}ski et al.}{2013}]{belczynski13} 
Belczy{\'n}ski K., Bulik T., Mandel I., Sathyaprakash B.~S., Zdziarski A.~A., Miko{\l}ajewska J., 2013, ApJ, 764, 96 

\bibitem[\protect\citeauthoryear{Blandford \& Konigl}{1979}]{bk79} 
Blandford R.~D., Konigl A., 1979, ApJ, 232, 34 

\bibitem[\protect\citeauthoryear{Bradt, Rothschild \& Swank}{Bradt et al.}{1993}]{brs93}
Bradt H. V., Rothschild R. E., Swank J. H., 1993, A\&AS, 97, 355

\bibitem[\protect\citeauthoryear{Bulgarelli et al.}{2012}]{bulgarelli12} 
Bulgarelli A., et al., 2012, A\&A, 538, A63

\bibitem[\protect\citeauthoryear{Burlon et al.}{2013}]{burlon13} 
Burlon D., Ghirlanda G., Murphy T., Chhetri R., Sadler E., Ajello M., 2013, MNRAS, 431, 2471
 
\bibitem[\protect\citeauthoryear{Casares \& Jonker}{2014}]{cj14} 
Casares J., Jonker P.~G., 2014, SSRv, 183, 223 

\bibitem[\protect\citeauthoryear{Chapuis \& Corbel}{2004}]{cc04} 
Chapuis C., Corbel S., 2004, A\&A, 414, 659 

\bibitem[\protect\citeauthoryear{Chu \& Bieging}{1973}]{cb73} 
Chu K. W., Bieging J. H., 1973, ApJ, 179, 21

\bibitem[\protect\citeauthoryear{Corbel et al.}{2000}]{corbel00} 
Corbel S., Fender R.~P., Tzioumis A.~K., Nowak M., McIntyre V., Durouchoux P., Sood R., 2000, A\&A, 359, 251 

\bibitem[\protect\citeauthoryear{Corbel et al.}{2003}]{corbel03} 
Corbel S., Nowak M.~A., Fender R.~P., Tzioumis A.~K., Markoff S., 2003, A\&A, 400, 1007 

\bibitem[\protect\citeauthoryear{Corbel et al.}{2004}]{corbel04} 
Corbel S., Fender R.~P., Tomsick J.~A., Tzioumis A.~K., Tingay S., 2004, 
ApJ, 617, 1272 

\bibitem[\protect\citeauthoryear{Corbel, Koerding \& Kaaret}{Corbel et al.}{2008}]{corbel08} 
Corbel S., Koerding E., Kaaret P., 2008, MNRAS, 389, 1697 

\bibitem[\protect\citeauthoryear{Corbel et al.}{2012}]{corbel12} 
Corbel S., et al., 2012, MNRAS, 421, 2947 

\bibitem[\protect\citeauthoryear{Corbel et al.}{2013a}]{corbel13a} 
Corbel S., Coriat M., Brocksopp C., Tzioumis A.~K., Fender R.~P., Tomsick J.~A., Buxton M.~M., Bailyn C.~D., 2013a, MNRAS, 428, 2500 

\bibitem[\protect\citeauthoryear{Corbel et al.}{2013b}]{corbel13b} 
Corbel S., et al., 2013b, MNRAS, 431, L107 

\bibitem[\protect\citeauthoryear{Coriat et al.}{2011}]{coriat11} 
Coriat M., et al., 2011, MNRAS, 414, 677 

\bibitem[\protect\citeauthoryear{Corral-Santana et al.}{2016}]{cs15} 
Corral-Santana J.~M., Casares J., Munoz-Darias T., Bauer F.~E., Martinez-Pais I.~G., Russell D.~M., 2016, A\&A, in press, arXiv:1510.08869 

\bibitem[\protect\citeauthoryear{Dickey}{1983}]{d83} 
Dickey J. M., 1983, ApJ, 273, L71

\bibitem[\protect\citeauthoryear{Dickey \& Lockman}{1990}]{dl90} 
Dickey J. M., Lockman F. J., 1990, ARA\&A, 28, 215

\bibitem[\protect\citeauthoryear{Done \& Gierli{\'n}ski}{2003}]{dg03} 
Done C., Gierli{\'n}ski M., 2003, MNRAS, 342, 1041 

\bibitem[\protect\citeauthoryear{Done, Wardzi{\'n}ski \& Gierli{\'n}ski}{Done et al.}{2004}]{dwg04} 
Done C., Wardzi{\'n}ski G., Gierli{\'n}ski M., 2004, MNRAS, 349, 393 

\bibitem[\protect\citeauthoryear{Done, Gierli{\'n}ski \& Kubota}{Done et al.}{2007}]{dgk07} 
Done C., Gierli{\'n}ski M., Kubota A., 2007, A\&ARv, 15, 1 

\bibitem[\protect\citeauthoryear{Ebisawa}{1998}]{ebisawa98} 
Ebisawa K., 1998, IAUS, 188, 392 

\bibitem[\protect\citeauthoryear{Fender, Hanson \& Pooley}{Fender et al.}{1999}]{fender99} 
Fender R.~P., Hanson M.~M., Pooley G.~G., 1999, MNRAS, 308, 473 

\bibitem[\protect\citeauthoryear{Fender et al.}{2000}]{fender00}
Fender, R. P., Pooley, G. G., Durouchoux, P., Tilanus, R. P. J., Brocksopp, C., 2000, MNRAS, 312, 853

\bibitem[\protect\citeauthoryear{Fermi LAT Collaboration}{2009}]{fermi09} 
Fermi LAT Collaboration, 2009, Sci, 326, 1512

\bibitem[\protect\citeauthoryear{Gallo, Fender \& Pooley}{Gallo et al.}{2003}]{gfp03}
Gallo E., Fender R.~P., Pooley G.~G., 2003, MNRAS, 344, 60

\bibitem[\protect\citeauthoryear{Giacconi et al.}{1967}]{giacconi67} 
Giacconi R., Gorenstein P., Gursky H., Waters J. R., 1967, ApJ, 148, L119

\bibitem[\protect\citeauthoryear{Gladstone, Done \& Gierli{\'n}ski}{Gladstone et al.}{2007}]{gdg07} 
Gladstone J., Done C., Gierli{\'n}ski M., 2007, MNRAS, 378, 13 

\bibitem[\protect\citeauthoryear{Harmon et al.}{2002}]{harmon02} 
Harmon B. A. et al., 2002, ApJS, 138, 149

\bibitem[\protect\citeauthoryear{Heinz \& Sunyaev}{2003}]{hs03} 
Heinz S., Sunyaev R.~A., 2003, MNRAS, 343, L59 

\bibitem[\protect\citeauthoryear{Hjalmarsdotter et al.}{2008}]{h08}
Hjalmarsdotter L., Zdziarski A. A., Larsson S., Beckmann V., McCollough M., Hannikainen D. C., Vilhu O., 2008, MNRAS, 384, 278

\bibitem[\protect\citeauthoryear{Hjalmarsdotter et al.}{2009}]{h09} 
Hjalmarsdotter L., Zdziarski A.~A., Szostek A., Hannikainen D.~C., 2009, MNRAS, 392, 251

\bibitem[\protect\citeauthoryear{in't Zand, Jonker \& Markwardt}{in't Zand et al.}{2007}]{ijm07} 
in't Zand J.~J.~M., Jonker P.~G., Markwardt C.~B., 2007, A\&A, 465, 953 

\bibitem[\protect\citeauthoryear{Koljonen et al.}{2010}]{koljonen10} 
Koljonen K.~I.~I., Hannikainen D.~C., McCollough M.~L., Pooley G.~G., Trushkin S.~A., 2010, MNRAS, 406, 307 

\bibitem[\protect\citeauthoryear{Krimm et al.}{2013}]{krimm13} 
Krimm H.~A., et al., 2013, ApJS, 209, 14 

\bibitem[\protect\citeauthoryear{Lauqu\'e et al.}{1972}]{laugue72} 
Lauqu\'e R., Lequeux J., Nguyen-Quang-Rieu, 1972, Nat.\ Phys.\ Sci., 239, 119

\bibitem[\protect\citeauthoryear{Laycock, Maccarone \& Christodoulou}{Laycock et al.}{2015}]{lmc15} 
Laycock, S. G. T., Maccarone T. J., Christodoulou D. M., 2015, MNRAS, 452, L31

\bibitem[\protect\citeauthoryear{Levine et al.}{1996}]{levine96}
Levine A. M., Bradt H., Cui W., Jernigan J. G., Morgan E. H.,
Remillard R., Shirey R. E., Smith D. A., 1996, ApJ, 469, L33

\bibitem[\protect\citeauthoryear{Ling, Zhang \& Tang}{Ling et al.}{2009}]{lzt09} 
Ling Z., Zhang S. \& Tang S., 2009, ApJ, 695, 1111

\bibitem[\protect\citeauthoryear{Loh et al.}{2016}]{fermi16} 
Loh A., Corbel S., Dubus G., Corbet R., 2016, ATel, 8591 

\bibitem[\protect\citeauthoryear{Markwardt et al.}{2005}]{m05} 
Markwardt C.~B., Tueller J., Skinner G.~K., Gehrels N., Barthelmy S.~D., Mushotzky R.~F., 2005, ApJ, 633, L77 

\bibitem[\protect\citeauthoryear{Mart{\'{\i}}, Paredes \& Peracaula}{Marti et al.}{2001}]{marti01} 
Mart{\'{\i}} J., Paredes J.~M., Peracaula M., 2001, A\&A, 375, 476 

\bibitem[\protect\citeauthoryear{Matsuoka et al.}{2009}]{matsuoka09} 
Matsuoka, M. et al., 2009, PASJ, 61, 999

\bibitem[\protect\citeauthoryear{McCollough et al.}{1999}]{mccollough99} 
McCollough M. L., et al.\ 1999, ApJ, 517, 951

\bibitem[\protect\citeauthoryear{Merloni, Heinz \& di Matteo}{Merloni et al.}{2003}]{mhd03} 
Merloni A., Heinz S., di Matteo T., 2003, MNRAS, 345, 1057 

\bibitem[\protect\citeauthoryear{Migliari \& Fender}{2006}]{mf06} 
Migliari S., Fender R.~P., 2006, MNRAS, 366, 79 

\bibitem[\protect\citeauthoryear{Mioduszewski et al.}{2001}]{m01} 
Mioduszewski A.~J., Rupen M.~P., Hjellming R.~M., Pooley G.~G., Waltman E.~B., 2001, ApJ, 553, 766 

\bibitem[\protect\citeauthoryear{Pooley \& Fender}{1997}]{pf97} 
Pooley G.~G., Fender R.~P., 1997, MNRAS, 292, 925

\bibitem[\protect\citeauthoryear{Poutanen, Zdziarski \& Ibragimov}{Poutanen et al.}{2008}]{pzi08} 
Poutanen J., Zdziarski A.~A., Ibragimov A., 2008, MNRAS, 389, 1427

\bibitem[\protect\citeauthoryear{Predehl et al.}{2000}]{p00}
Predehl P., Burwitz V., Paerels F., Tr{\"u}mper J., 2000, A\&A, 357, L25 

\bibitem[\protect\citeauthoryear{Reeves et al.}{2008}]{reeves08} 
Reeves J., Done C., Pounds K., Terashima Y., Hayashida K., Anabuki N., Uchino M., Turner M., 2008, MNRAS, 385, L108 

\bibitem[\protect\citeauthoryear{Reid et al.}{2014}]{reid14} 
Reid M.~J., McClintock J.~E., Steiner J.~F., Steeghs D., Remillard R.~A., Dhawan V., Narayan R., 2014, ApJ, 796, 2 

\bibitem[\protect\citeauthoryear{Remillard \& McClintock}{2006}]{rm06} 
Remillard R.~A., McClintock J.~E., 2006, ARA\&A, 44, 49 

\bibitem[\protect\citeauthoryear{Rushton et al.}{2010}]{rushton10} 
Rushton A., Spencer R., Fender R., Pooley G., 2010, A\&A, 524, A29 

\bibitem[\protect\citeauthoryear{Salvato, Greiner \& Kuhlbrodt}{Salvato et al.}{2004}]{salvato04} 
Salvato M., Greiner J., Kuhlbrodt B., 2004, ApJ, 600, L31 

\bibitem[\protect\citeauthoryear{Segreto et al.}{2010}]{segreto10} 
Segreto A., Cusumano G., Ferrigno C., La Parola V., Mangano V., Mineo T.,  Romano P., 2010, A\&A, 510, A47

\bibitem[\protect\citeauthoryear{Szostek \& Zdziarski}{2008}]{sz08} 
Szostek A., Zdziarski A. A., 2008, MNRAS, 386, 593

\bibitem[\protect\citeauthoryear{Szostek, Zdziarski \& McCollough}{Szostek et al.}{2008}]{szm08} 
Szostek A., Zdziarski A.~A., McCollough M.~L., 2008, MNRAS, 388, 1001 (SZM08)

\bibitem[\protect\citeauthoryear{Tavani et al.}{2009}]{agile} 
Tavani M., et al., 2009, Nature, 462, 620 

\bibitem[\protect\citeauthoryear{Tavani et al.}{2016}]{agile16} 
Tavani M., et al., 2016, ATel, 8597

\bibitem[\protect\citeauthoryear{Uttley, McHardy \& Vaughan}{Uttley et al.}{2005}]{uttley05} 
Uttley P., McHardy I.~M., Vaughan S., 2005, MNRAS, 359, 345 

\bibitem[\protect\citeauthoryear{van Kerkwijk et al.}{1992}]{v92} 
van Kerkwijk M.~H., et al., 1992, Nature, 355, 703 

\bibitem[\protect\citeauthoryear{van Kerkwijk}{1993}]{v93} 
van Kerkwijk M.~H., 1993, A\&A, 276, L9 

\bibitem[\protect\citeauthoryear{van Kerkwijk et al.}{1996}]{v96} 
van Kerkwijk M. H., Geballe T. R., King D. L., van der Klis M.,
van Paradijs J., 1996, A\&A, 314, 521

\bibitem[\protect\citeauthoryear{Waltman et al.}{1994}]{waltman94} 
Waltman E.~B., Fiedler R.~L., Johnston 
K.~J., Ghigo F.~D., 1994, AJ, 108, 179 

\bibitem[\protect\citeauthoryear{Waltman et al.}{1995}]{waltman95} 
Waltman E.~B., Ghigo F.~D., Johnston K.~J., Foster R.~S., Fiedler R.~L., Spencer J.~H., 1995, AJ, 110, 290 

\bibitem[\protect\citeauthoryear{Waltman et al.}{1996}]{waltman96} 
Waltman E. B., Foster R. S., Pooley G. G., Fender R. P., Ghigo F. D., 1996, AJ, 112, 2690

\bibitem[\protect\citeauthoryear{Watanabe et al.}{1994}]{watanabe94} 
Watanabe H., Kitamoto S., Miyamoto S., Fielder R. L., Waltman E. B., Johnston K. J., Ghigo F. D., 1994, ApJ, 433, 350

\bibitem[\protect\citeauthoryear{Williams et al.}{2011}]{williams11} 
Williams P.~K.~G., et al., 2011, ApJ, 733, L20 

\bibitem[\protect\citeauthoryear{Yoon \& Heinz}{2015}]{yh15}
Yoon D., Heinz S., 2015, ApJ, 801, 55

\bibitem[\protect\citeauthoryear{Yoon, Zdziarski \& Heinz}{Yoon et al.}{2016}]{zyh15} 
Yoon D., Zdziarski A.~A., Heinz S., 2016, MNRAS, 456, 3638

\bibitem[\protect\citeauthoryear{Zdziarski}{2012}]{zdz12}
Zdziarski A. A., 2012, MNRAS, 422, 1750

\bibitem[\protect\citeauthoryear{Zdziarski et al.}{2002}]{z02} 
Zdziarski A.~A., Poutanen J., Paciesas W.~S., Wen L., 2002, ApJ, 578, 357 

\bibitem[\protect\citeauthoryear{Zdziarski et al.}{2004}]{z04}
Zdziarski A. A., Gierli\'nski M., Miko{\l}ajewska J., Wardzi\'nski G., Smith D. M., Harmon B. A., Kitamoto S., 2004, MNRAS, 351, 791

\bibitem[\protect\citeauthoryear{Zdziarski, Misra \& Gierli{\'n}ski}{Zdziarski et al.}{2010}]{zmg10} 
Zdziarski A.~A., Misra R., Gierli{\'n}ski M., 2010, MNRAS, 402, 767 

\bibitem[\protect\citeauthoryear{Zdziarski et al.}{2011}]{zspl11} 
Zdziarski A.~A., Skinner G.~K., Pooley G.~G., Lubi{\'n}ski P., 2011, MNRAS, 416, 1324 (ZSPL11)

\bibitem[\protect\citeauthoryear{Zdziarski et al.}{2012}]{z12b} 
Zdziarski A.~A., Maitra C., Frankowski A., Skinner G.~K., Misra R., 2012, MNRAS, 426, 1031

\bibitem[\protect\citeauthoryear{Zdziarski, Miko{\l}ajewska \& Belczy{\'n}ski}{Zdziarski et al.}{2013}]{zmb13} 
Zdziarski A.~A., Miko{\l}ajewska J., Belczy{\'n}ski K., 2013, MNRAS, 429, L104 

\end{thebibliography}
\end{document}